\documentclass{jfm}
\input{amssym.def}
\input{amssym.tex}
\usepackage{amsmath}
\usepackage{todonotes}
\usepackage{natbib}
\usepackage{psfrag} 
\usepackage{graphicx} 
\usepackage{xcolor}
\usepackage{subfigure}
\usepackage{relsize}
\def\drawline#1#2{\raise 2.5pt\vbox{\hrule width #1pt height #2pt}}
\def\spacce#1{\hskip #1pt}
\def\solid{\drawline{24}{.5}\nobreak}

\def\bdash{\hbox{\drawline{4}{.5}\spacce{2}}}
\def\bdashb{\hbox{\drawline{3}{.5}\spacce{1}}}
\def\bdashc{\hbox{\drawline{5}{.5}\spacce{2}}}
\def\dashed{\bdash\bdash\bdash\bdash\nobreak}
\def\dashedb{\bdashb\bdashb\spacce{3}\bdashb\bdashb\nobreak}
\def\bdot{\hbox{\drawline{0.5}{.5}\spacce{2}}}
\def\dashdot{\bdashc\bdot\bdashc\bdot\bdashc\nobreak}

\def\chndot{\hbox {\drawline{9.5}{.5}\spacce{2}\drawline{1}{.5}\spacce{2}\drawline{9.5}{.5}}\nobreak\ } 
 
\def\trian{\raise 1.25pt\hbox{$\scriptscriptstyle\triangle$}\nobreak}
\def\tridot{$\triangle$\hspace{-0.6em}$\cdot$\hspace{0.4em}\nobreak}

\def\square{${\vcenter{\hrule height .4pt
        \hbox{\vrule width .4pt height 3pt \kern 3pt
        \vrule width .4pt}
        \hrule height .4pt}}$\nobreak\ }

\definecolor{salmon}{RGB}{255,76,0}

\begin{document}
\newcommand{\pz}{\pagebreak[0]}

\title[
Comparison between SHS,
LIS and rough surfaces]
{
Comparison between super-hydrophobic,
liquid infused and rough surfaces: a DNS study.
}


\author[ I. Arenas, E. Garcia, M. K. Fu, P. Orlandi, M. Hultmark  and S. Leonardi]
{Isnardo Arenas$^{\rm a}$, Edgardo Garc\'ia$^{\rm b}$,
Matthew K. Fu$^{\rm c}$, 
Paolo Orlandi$^{\rm d}$, 
Marcus Hultmark$^{\rm c}$ and Stefano Leonardi$^{\rm b}$ \thanks{Corresponding author. Email: stefano.leonardi@utdallas.edu}}

\affiliation{
$^{\rm a}${\em{Dep. de ciencias b\'asicas, Unidades Tecnol\'ogicas de Santander; Bucaramanga, Colombia}}; \\
$^{\rm b}${\em{Dept. of Mechanical Engineering, The University of Texas at Dallas}}; \\
$^{\rm c}${\em{Dept. of Mechanical and Aerospace Engineering, Princeton University}};  \\
$^{\rm d}${\em{Dipartimento di Meccanica ed Aeronautica, Universit\'a di Roma La Sapienza}};  
}

\date{\today}

\maketitle

\begin{abstract}
	Direct Numerical Simulations of two superposed fluids in a channel with a textured surface on the lower wall have been carried out. A parametric study varying the viscosity ratio between the two fluids has been performed to mimic both { idealised} super--hydrophobic and liquid--infused surfaces and assess its effect on the frictional, form and total drag for three different textured geometries: longitudinal square bars, transversal square bars and staggered cubes. The interface between the two fluids is assumed to be slippery in the streamwise and spanwise directions and not deformable in the vertical direction, corresponding to the ideal case of infinite surface tension. 
To identify the role of the fluid-fluid interface, an extra set of simulations with a single fluid has been 
carried out. Comparison with the cases with two fluids reveals the role of the interface in suppressing turbulent transport between the lubricating layer and the overlying flow decreasing the overall drag.
In addition, the drag and the maximum wall--normal velocity fluctuations were found to be highly correlated 
	for all the surface configurations, whether they reduce or increase the drag. This implies that the structure of the near--wall turbulence is  dominated by the total shear and not by the local boundary condition of super--hydrophobic, liquid--infused or rough surfaces.
\end{abstract}

\section{Introduction}
There have been considerable advancements in the design and fabrication of surface treatments that can passively reduce turbulent skin friction drag. Often this is accomplished with microscale features that are designed to interact with the smallest (viscous) scales of the turbulence and modify the near wall flow. A particular example of this is the use of riblets, which rely on streamwise protrusions to reduce interactions between the overlying turbulence and solid surface \citep{Dean2010a,Mayoral2011}. Tailoring the scales and topography of riblets allows them to exhibit behavior distinct from that of conventional roughness in a particular range of flow conditions. However, outside of this operating regime, the beneficial effects of the surface configuration will be negated and likely generate increased drag. 

Recently, there has been significant interest in superposing a second fluid onto the roughness to improve the drag reduction characteristics. In particular, super--hydrophobic surface [SHS] have garnered the bulk of this interest and are an extremely promising method of passive drag reduction. These surfaces are composed of hydrophobic surface textures which can locally stabilize pockets of air when submerged underwater, resulting in a heterogeneous surface of air-water and solid-water interfaces. The slip effect facilitated by the presence of these air-water interfaces has been demonstrated to reduce turbulent drag in wide array of experimental facilities \citep{Daniello2009,Ling2016,Srinivasan2015,Gose2018} at magnitudes up to $75\%$ \citep{Park2014}. 
Theoretical and computational studies of turbulent flow over SHS have demonstrated similarly impressive results \citep{Min2004,Martell2010,Fukagata2006b,Turk2014,Jelly2014,Seo2015a,Seo2016,Rastegari2015}, though the details of the flow within the air layer have often been neglected by modeling the surface boundary condition as a pattern of shear-free and no-slip boundaries. Such an assumption has been justified by the large dynamic viscosity ratio between air and water and only recently have studies (\cite{Jung2016}, \cite{Li2017}) started to consider the influence of the flow within the air layer. Despite the improvements to our understanding of the interactions between SHS and turbulence, there are still many open questions regarding the influence of surface morphology on the overlying turbulence and drag reduction mechanism. 

A surface treatment similar to SHS is liquid-infused surfaces [LIS]. Inspired by the Nepenthes pitcher plant \citep{Wong2011}, LIS are composed of functionalized surface textures wetted with an immiscible, chemically-matched liquid lubricant. Like SHS, the resulting surface is heterogeneous, containing a mixture of fluid-liquid and fluid-solid interfaces. In addition to exhibiting a wide range of desirable properties including ice-phobicity \citep{Epstein2012}, pressure-stability, self-cleaning, and omniphobicity \citep{Wong2011}, it has been experimentally demonstrated that grooved LIS configurations can reduce turbulent drag up to $35\%$ \citep{Rosenberg2016,VanBuren2017}. 
Furthermore, from direct numerical simulations [DNS] results,  \citet{Fu2017JFM} found that the drag reduction mechanism exhibited by LIS is fundamentally the same as SHS. 
{ This was further corroborated by \citet{Rastegari2019}. 
They correlated the amount of drag reduction obtained with 
SHS and LIS with the shift ($B-B_0$) in the intercept of the 
logarithmic law of the wall relative to the baseline smooth wall.
The relation between drag reduction 
and $B-B_0$ was shown to be the same for SHS and LIS.
}
However, in the case of LIS, the viscosity of the lubricating fluid is of the same order  as the overlying fluid and plays an important role in determining whether a LIS will reduce or increase drag. This is distinct from the case of SHS, where it has typically been assumed that the drag reduction is inherently due to the negligible viscosity of the air layer compared to the external fluid. In contrast, recent results from several LIS studies \citep{Fu2017JFM,Rosenberg2016,VanBuren2017} have found that LIS can generate turbulent drag reduction even when the lubricant is more viscous than the external fluid. However, when the lubricant viscosity significantly exceeded that of the external fluid the surfaces were found to increase drag. While LIS can be analyzed and considered within the same frame as SHS, many of the same questions regarding the influence of surface morphology, and lubricant properties, on the resulting drag reduction remain open.

Here, we study the drag behavior of canonical, structured surfaces superposed with a second fluid. Specifically, we detail how the viscosity of the superposed fluid and surface morphology influence the overall drag budget, i.e. the relative contributions of form and friction drag, and the role of the fluid-fluid interface in suppressing turbulent transport between the lubricating layer and overlying flow. To quantify these contributions, we performed DNS of a turbulent channel flow with the lower wall consisting of either longitudinal, transversal square bars or staggered cubes wetted with a second fluid. The viscosity of the second fluid is varied over a wide range to simulate both idealized SHS and LIS and the results were compared to a smooth wall DNS with the same mass flux. To identify the role of the fluid-fluid interface, the results were compared to a series of simulations with the same surface features but only a single fluid (i.e. rough wall). Furthermore, by considering a wide array of surface configurations we attempt to identify some common  flow behaviors to SHS, LIS and rough surfaces.

\section{Flow configuration}

\begin{figure}
\begin{center}
\psfrag{kp}{$k$}
\psfrag{lp}{$l$}
\psfrag{sp}{$p$}
\psfrag{sp1}{$p$}
\psfrag{sp2}{$p$}
\psfrag{a}{a)}
\psfrag{b}{b)}
\psfrag{c}{c)}
\psfrag{d}{d)}
\psfrag{e}{e)}
\psfrag{f}{f)}
\psfrag{x1}{\relsize{+2} $x$}
\psfrag{x2}{\relsize{+2} \hskip 0.4cm $y$}
\psfrag{x3}{\relsize{+2} \hskip -0.1cm $z$}
\psfrag{c1}{\relsize{+2} $k$}
\psfrag{c2}{\relsize{+2} $w$}
\psfrag{c3}{\relsize{+2} $k$}
\psfrag{c4}{\relsize{+2} $h$}
\psfrag{c5}{\relsize{+2} $h$}
\psfrag{c6}{\relsize{+2} \hskip -1cm flow dir.}
a)
\resizebox*{0.45\textwidth}{!}{\includegraphics{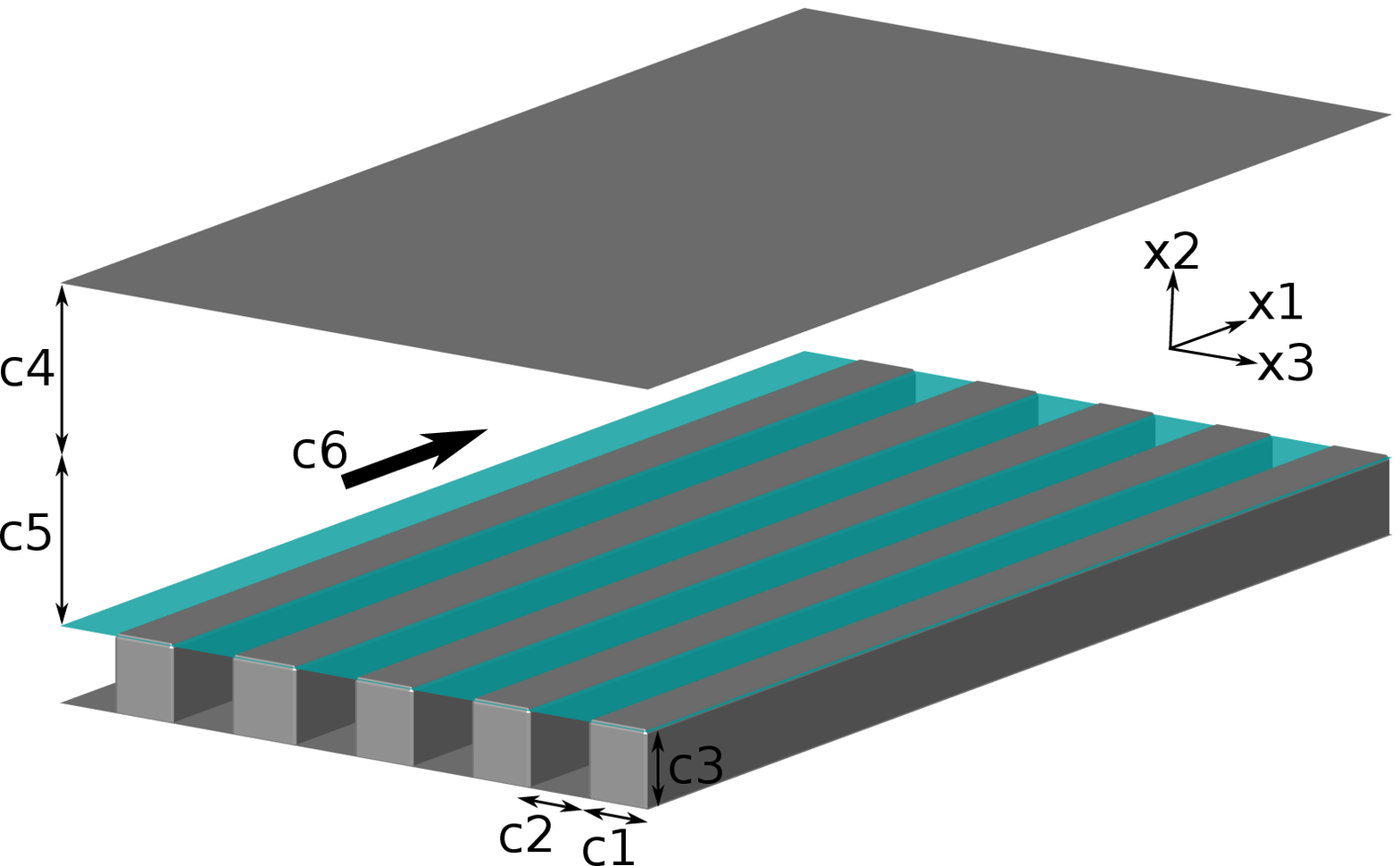} } 
b)
\resizebox*{0.45\textwidth}{!}{\includegraphics{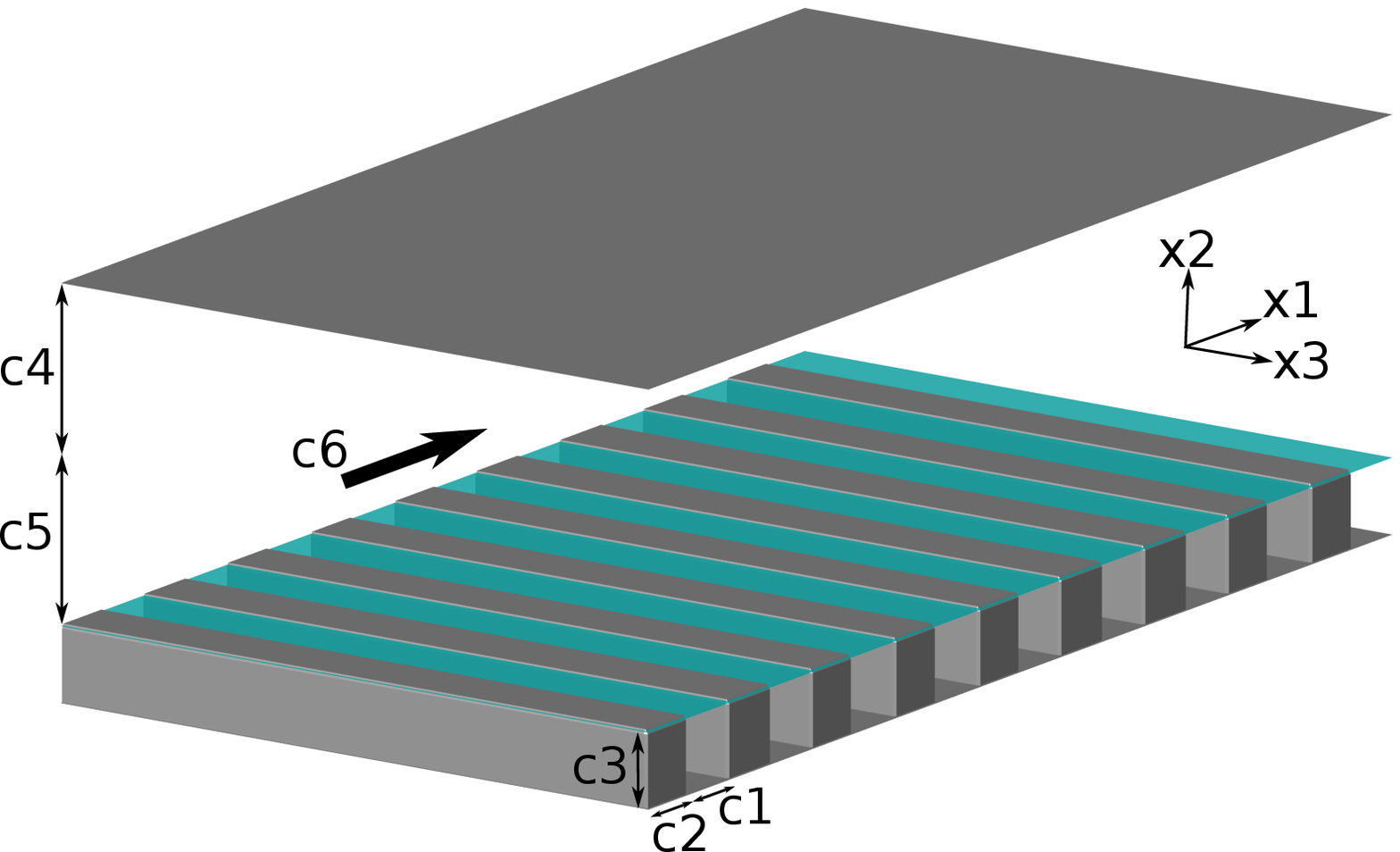} } \\
c)
\resizebox*{0.45\textwidth}{!}{\includegraphics{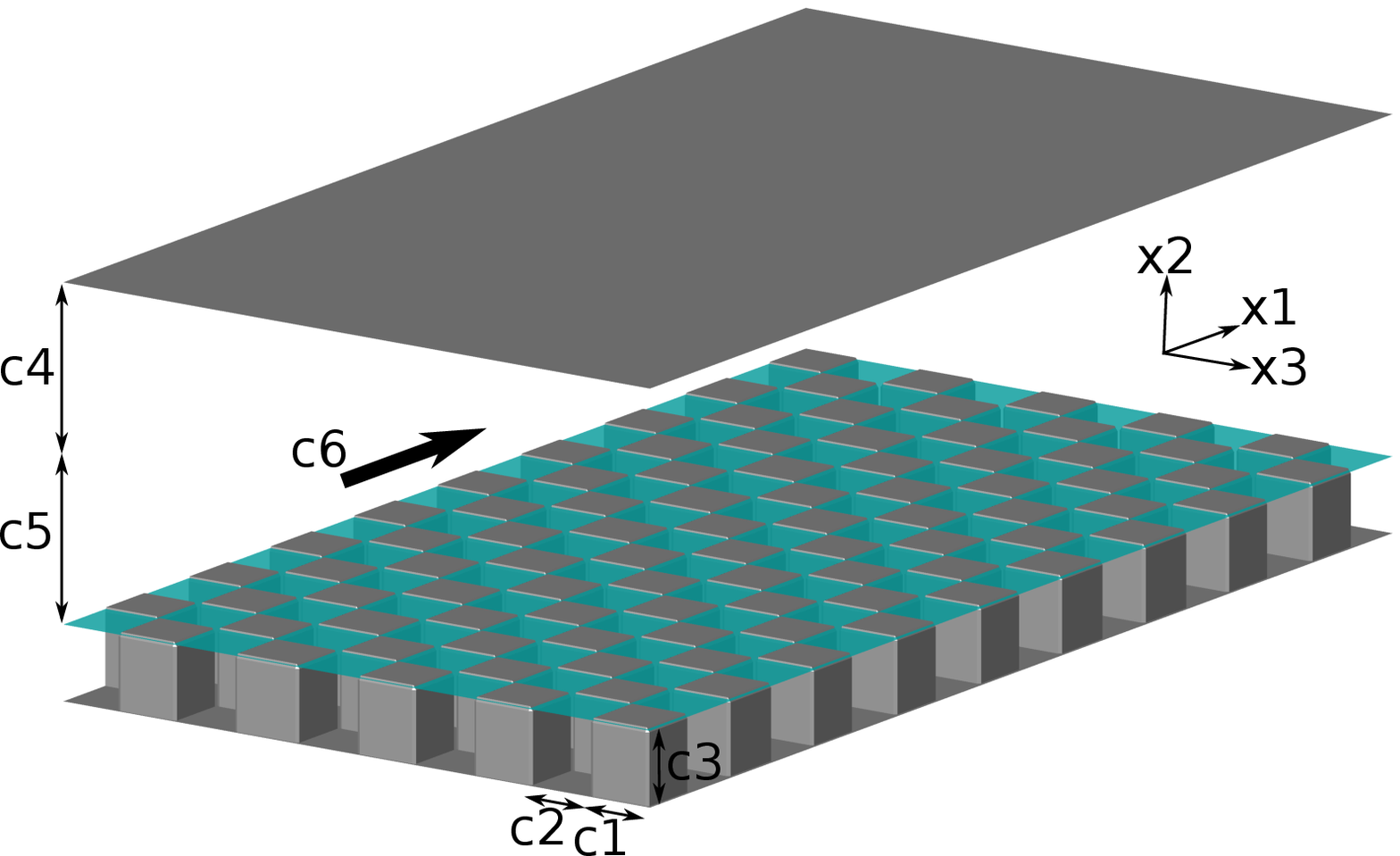} }
d)
\resizebox*{0.45\textwidth}{!}{\includegraphics{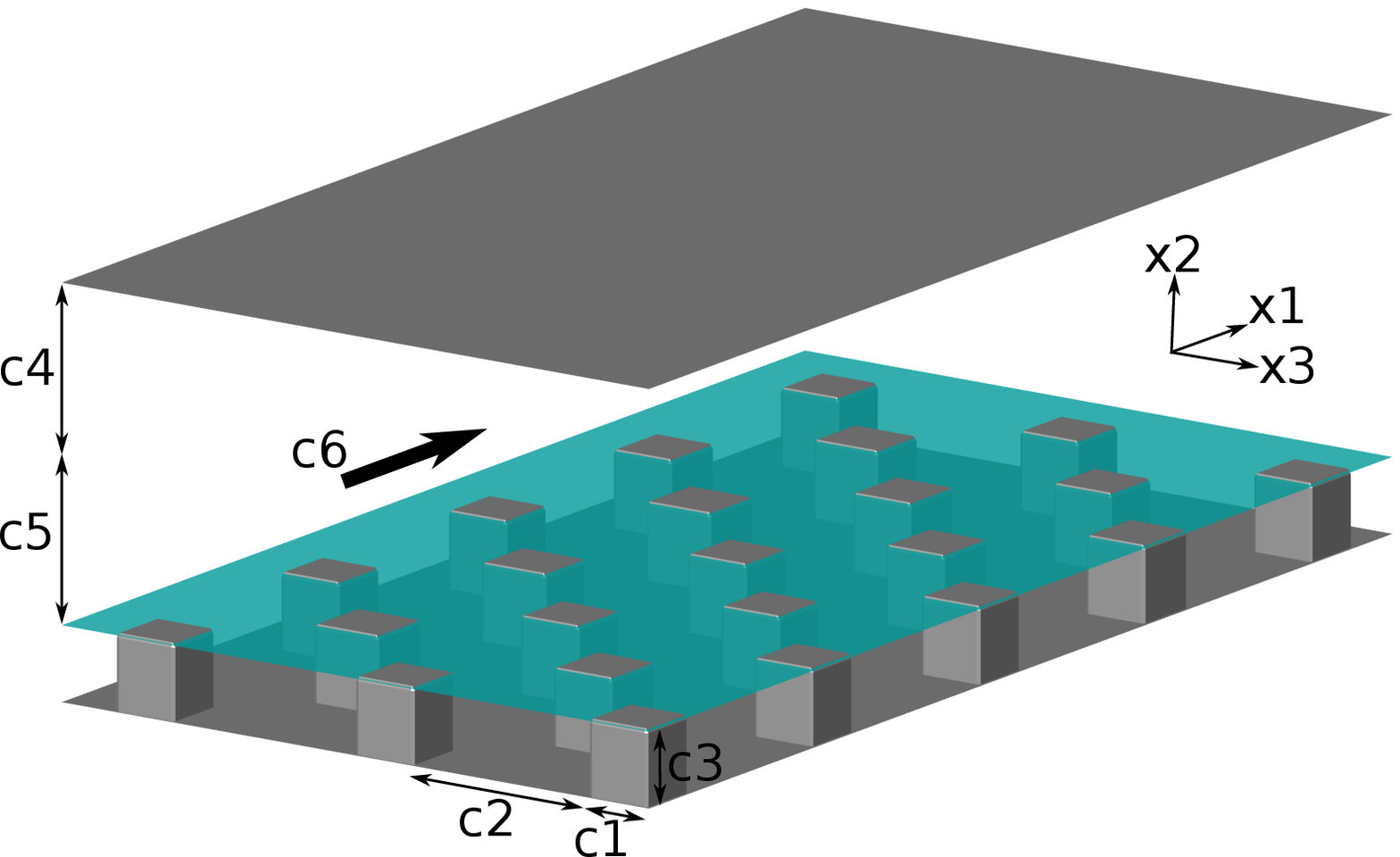} } \\
\caption{Geometrical sketch of the textured surfaces:
a) Longitudinal square bars (LSB) with $a=0.5$, b) transversal square bars (TSB) with $a=0.5$,  
staggered cubes (SC) c) $a=0.5$ and d) $a=0.875$. The interface between the
	two fluids is the horizontal surface at the crests plane colored in turquoise. 
	{	The dimension of the texture $k$ is not to scale with $h$ for presentation purposes}.}
\label{fig:domain}
\end{center}
\end{figure}
The upper wall of the channel is smooth while
the lower wall is  made of either
longitudinal, transversal square bars or staggered cubes (see Figure \ref{fig:domain}).
The interface between the two fluids is placed at the crests plane (turquoise surface in the figure), separating the main stream
from the fluid in the cavities. 
To minimize the number of parameters, we focus our attention on the role of the lubricant viscosity, specifically the viscosity ratio $N$, where $N={\mu_2}/{\mu_1}$ (where the subscript $1$ and $2$ indicate the fluid in and above the cavities, respectively). Several viscosity ratios between the two fluids have been considered, 
ranging from $N=0.1$ to $N=100$. 
The density is assumed to be the same in both fluids and the buoyancy is neglected. 
High values of $N$ mimic 
the { the viscosity ratio} over SHS while low values of $N$ represent LIS. { The present paper does not attempt to reproduce a real SHS and LIS, but focus on 
an idealised model in order to understand the mechanism leading to drag reduction. The deformation of the interface, the dynamics of the contact line, the fluid retention and 
the effect of the density gradient at the interface are not addressed in the present paper.}

The origin in the vertical direction is at the
centerline, so the upper wall is at $y/h=1$ and the interface  at $y/h=-1$
(where $h$ is the half height of the channel).
The computational box is $6.4h \times 2.05h \times 3.2 h$ in
$x_1$ (streamwise), $x_2$ (wall-normal) and $x_3$ (spanwise direction),
respectively.
The additional $k=0.05h$ increase in  channel height corresponds to the cavity
height of the textured surfaces.

When both walls are smooth the bulk Reynolds number is $Re = U_b h/\nu_2=2,800$ and 
the turbulent Reynolds number is
$Re_{\tau} =U_\tau h/\nu_2 \simeq 177$  
(with $U_b$ bulk velocity and $U_\tau$ the friction 
velocity, and $\nu_2=\mu_2/\rho$).
The height of the cavities in the substrate in wall units is approximately 
$k^+\simeq 9$, "+" denotes normalization by
 $\nu/U_{\tau}$, where
 $U_{\tau} = (\tau/\rho)^{1/2}$ and $\tau$ is the wall shear.
Periodic boundary conditions were applied in the streamwise and spanwise directions
while the no-slip conditions are imposed on both the smooth (upper) and textured (lower) walls.

{ Details of the grid, turbulent Reynolds number and
a grid sensitivity study are discussed in the appendix}.
For both transversal square bars and longitudinal square bars, the pitch to width ratio is $p/w=2$,  ($p=w+k$),
corresponding to fluid area fraction, $a=w/p=0.5$ (i.e. the fraction of fluid-fluid area over the total interfacial area of the substrate).
Two different configurations of staggered cubes were evaluated, one with the same fluid-area fraction ($a=0.5$) and another with a larger fluid-area fraction ($a=0.875$). { The larger gas fraction is aimed at highlighting the role of the interface. In fact,
it should be expected that the surface with large $a$ would generate high values of drag 
reduction for SHS/LIS \citep{Fu2017JFM} but increase the drag when 
there is only one fluid in the channel, i.e. the case of a classical rough wall \citep{Leonardi2010}. }

\section{Numerical Procedure}
The flow in the domain was computed using the non-dimensional, incompressible Navier-Stokes and continuity
equations given by

\begin{eqnarray}\label{eq:NS-CM}
  & & \frac{\partial U_i}{\partial t}+\frac{\partial U_i U_j}{\partial x_j} = -
  \frac{\partial P}{\partial x_i} + \frac{1}{Re} \frac{\partial}{\partial x_j} \left [ \widetilde{\mu} (y/h)
  \left ( \frac{\partial U_i}{\partial x_j} + \frac{\partial U_j}{\partial x_i}\right ) \right ] +
  \Pi  \delta_{i1} \;, \\
  & & \frac{\partial U_i}{\partial x_i}=0 \;,
\end{eqnarray}
where  $U_{i}$ is the component of the velocity vector in the $i$
direction, $i=1$ is for the component $U$ in streamwise direction ($x$), 
$i=2$ is for the component $V$ in wall normal direction ($y$), 
and $i=3$ is for the component $W$ in spanwise direction ($z$), $\Pi$ is the pressure gradient required to maintain a constant
flow rate
and $P$ the pressure. 
The bulk Reynolds number is defined using the viscosity of the fluid in the main channel, $Re = \rho U_b h / \mu_2$. 
The change of viscosity in the cavities is accounted for with the term $\widetilde{\mu}$ using the
Heaviside function $H$,
defined as
\begin{equation}
  \widetilde{\mu}  (y/h) = \frac{1}{N} + \left(1 - \frac{1}{N}\right) H (y/h) \;,
\end{equation}
where $N$ is the viscosity ratio.
{  
In the present paper, the position of the interface is fixed at $y/h=-1$.
 For  $y/h<-1$, fluid  in the cavities (fluid 1),
 $H=0$, for $y/h>-1$, fluid above the cavities (fluid 2), $H=1$.

The wall normal velocity at the interface is zero and the shear stress is continuous:
 $\mu \partial U_1/\partial y$, $\mu \partial U_3/\partial y$. This mimics
 a stable interface and flow in Cassie state corresponding to {an infinite} surface tension.
The interface is, therefore, slippery in the spanwise and streamwise directions
but cannot be deformed in the vertical direction. 
}
The flow rate has been kept constant in all simulations.

Equations 3.1 and 3.2 
were discretized in an orthogonal coordinate system using a
staggered, central, second-order finite-difference approximation.  Additional details of the numerical
method can be found in \citet{Orlandi2000}.  The discrete system is
advanced in time using a fractional-step method with viscous terms
treated implicitly and convective terms explicitly.  The large sparse
matrix resulting from the implicit terms is inverted by an approximate
factorization technique.  At each time step, the momentum equations are
advanced with the pressure at the previous step, yielding an
intermediate non-solenoidal velocity field.  A scalar quantity $\Phi$
projects the non-solenoidal field onto a solenoidal one.  A hybrid
low-storage third-order Runge-Kutta scheme is used to advance the
equations in time.  The shape of the substrate is treated by the efficient immersed
boundary technique described in detail by \citet{Orlandi2006}.
This approach allows the solution of flows over complex geometries
without the need of computationally intensive body-fitted grids.  It
consists of imposing $U_{i}=0$ on the points occupied by the solid texture.
{ The discretization needs to be modified on the boundary cells otherwise the texture 
would be described in a step-wise way.
At the closest grid points to the boundary, the derivatives in the
Navier-Stokes equations are discretized using the distance between
the velocities and the boundary of the texture rather than using the mesh size.
}

 Statistics are computed with about $300$ velocity 
fields, $2$ non-dimensional time units apart (time is normalized by $h/U_b$).
After the first $3000$ non-dimensional time units, which were discarded,
 convergence to a statistically steady state was achieved.

{ 
The instantaneous fields (e.g. velocity or pressure) can be expressed as the
superposition of three components (\cite{Hussain1970}, \cite{Raupach1982}).
Considering for example the generic velocity component in the $i$ direction, $U_i(x,y,z,t)$, (the same notation applies to any other variable),
the instantaneous quantity
 can be decomposed as follows
\begin{equation}
U_i(x_l,y_m,z_k,t_n)=\overline{U_i}(y_m)+\tilde{U_i}(x_l,y_m,z_k)+u'_i(x_l,y_m,z_k,t_n),
\end{equation}
where $l,m,k$ are the indices of the grid in $x,y,z$ respectively and the index $n$ is for time.
An overline indicates averaging with respect to time, spanwise and streamwise directions:
\begin{equation}
	\overline{U_i}(y_m)=\frac{1}{N_x  N_z N_t}  \sum^{N_x}_{l=1}  \sum_{k=1}^{N_z} \sum_{n=1}^{N_t} U_i(x_l,y_m,z_k,t_n),
\end{equation}
where $N_t$ is the number of fields saved in
time, $N_z$ the points in the spanwise direction,  and $N_x$ the points in the streamwise direction, 
the grid is uniform in $x$ and $z$.
Angular brackets indicate averages with respect to time:
\begin{equation}
	\langle {U_i}(x_l,y_m,z_k)\rangle=\overline{U_i}(y_m)+\tilde{U_i}(x_l,y_m,z_k)=\frac{1}{ N_t}  \sum_{n=1}^{N_t} U_i(x_l,y_m,z_k,t_n)   \; .
\end{equation}
The Reynolds stresses are
\begin{equation}
	\overline{u_i u_j}(y_m)= \frac{1}{N_x  N_z N_t}  \sum_{l=1}^{N_x}  \sum_{k=1}^{N_z} \sum_{n=1}^{N_t} (U_i(x_l,y_m,z_k,t_n) - \overline{U_i}(y_m)) (U_j(x_l,y_m,z_k,t_n) - \overline{U_j}(y_m))  \;.
\end{equation}
They can be decomposed as the sum of a dispersive  component and incoherent component
substituting $U(x_l,y_m,z_k,t_n)-\overline{U}(y_m)=\tilde{U}(x_l,y_m,z_k)+u'(x_l,y_m,z_k,t_n)$:
\begin{equation}
  \begin{aligned}
	\overline{u_i u_j}(y_m)= \frac{1}{N_x  N_z N_t} \sum_{l=1}^{N_x}  \sum_{k=1}^{N_z} \sum_{n=1}^{N_t} 
	   (\tilde{U}_i(x_l,y_m,z_k)+u_i'(x_l,y_m,z_k,t_n)) & (\tilde{U_j}(x_l,y_m,z_k)+ \\
	                                                	 +u_j'(x_l,y_m,z_k,t_n)) \;,
  \end{aligned}
	\label{eq:Restress1}
\end{equation}
and then 
\begin{equation}
  \begin{aligned}
	\underbrace{\overline{u_i u_j}(y_m)}_\text{Reynolds stress}=
	  \frac{1}{N_x  N_z N_t}  \sum_{l=1}^{N_x}  \sum_{k=1}^{N_z} \sum_{n=1}^{N_t} 
	  &	\underbrace{\tilde{U}_i(x_l,y_m,z_k)\tilde{U}_j(x_l,y_m,z_k)}_\text{Dispersive component} +\\ 
	  & \underbrace{u_i'(x_l,y_m,z_k,t_n) u_j'(x_l,y_m,z_k,t_n)}_\text{ Incoherent component}  \;.
  \end{aligned}
\label{eq:Restress2}
\end{equation}
The first term on the right hand side is the dispersive stress which arises as a consequence of 
 the  spatial inhomogeneities in the mean flow and the second term is the incoherent component (the mixed product, $\tilde{U}_i(x_l,y_m,z_k)u_i'(x_l,y_m,z_k,t_n)$, once averaged in time is zero).
The relative contribution of the dispersive stresses to the total stress is discussed in Section 6.
Below the roughness crest, averages in space are carried out on the fluid portion 
only (the sum is divided by the area of the fluid and not by the total horizontal
area which includes the solid).}

\section{Mean Velocity}

\begin{figure}
\begin{center}
\psfrag{xlab}{\relsize{+2} \hspace{-1em}$\langle U \rangle/U_b$}
\psfrag{ylab}{\relsize{+2} $y/h$}
\psfrag{xlab1}{\relsize{+2} $z/p$}
\psfrag{ylab2}{\relsize{+2} $U_s/U_b$}
\psfrag{lam}{$\lambda$}
\psfrag{P1}{$P_1$}
\psfrag{P2}{$P_2$}
\psfrag{P3}{$P_3$}
a)
	\resizebox*{0.45\textwidth}{!}{\includegraphics{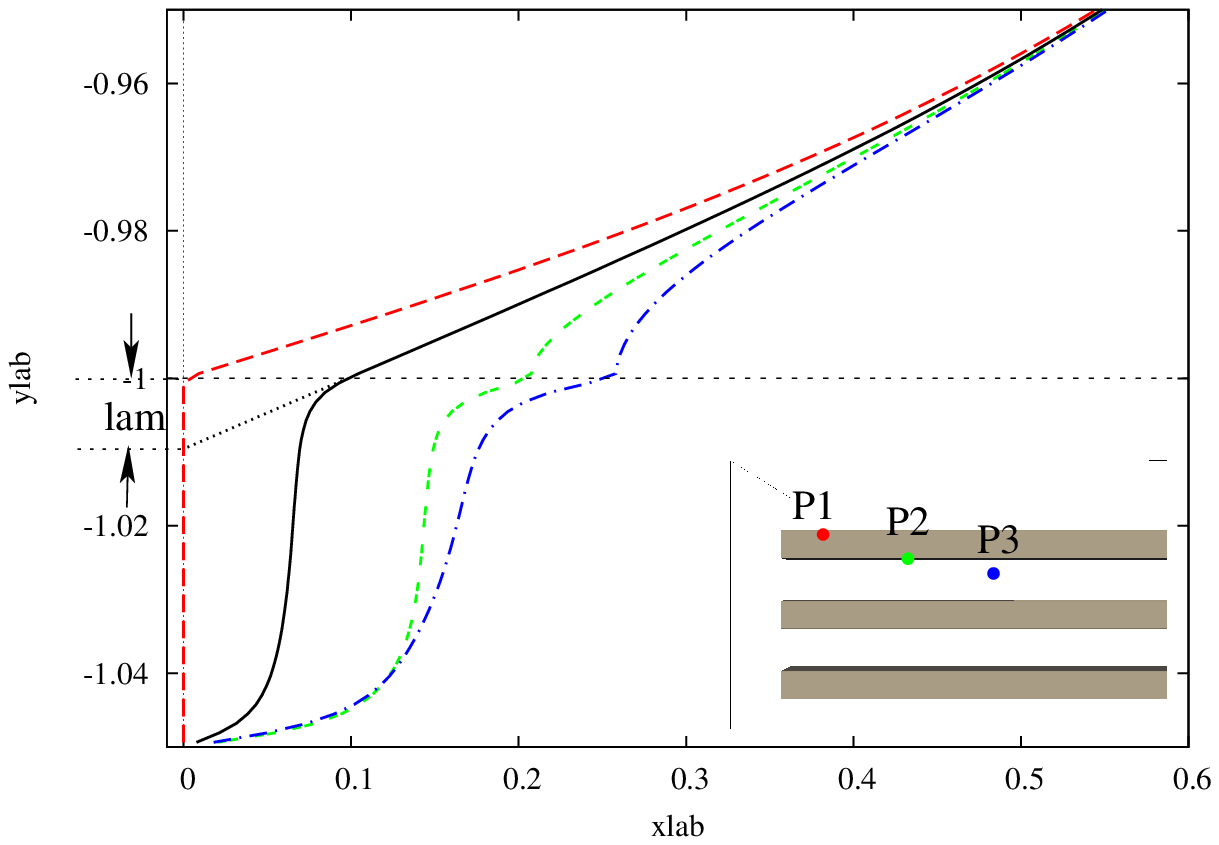}} 
b)
	\resizebox*{0.45\textwidth}{!}{\includegraphics{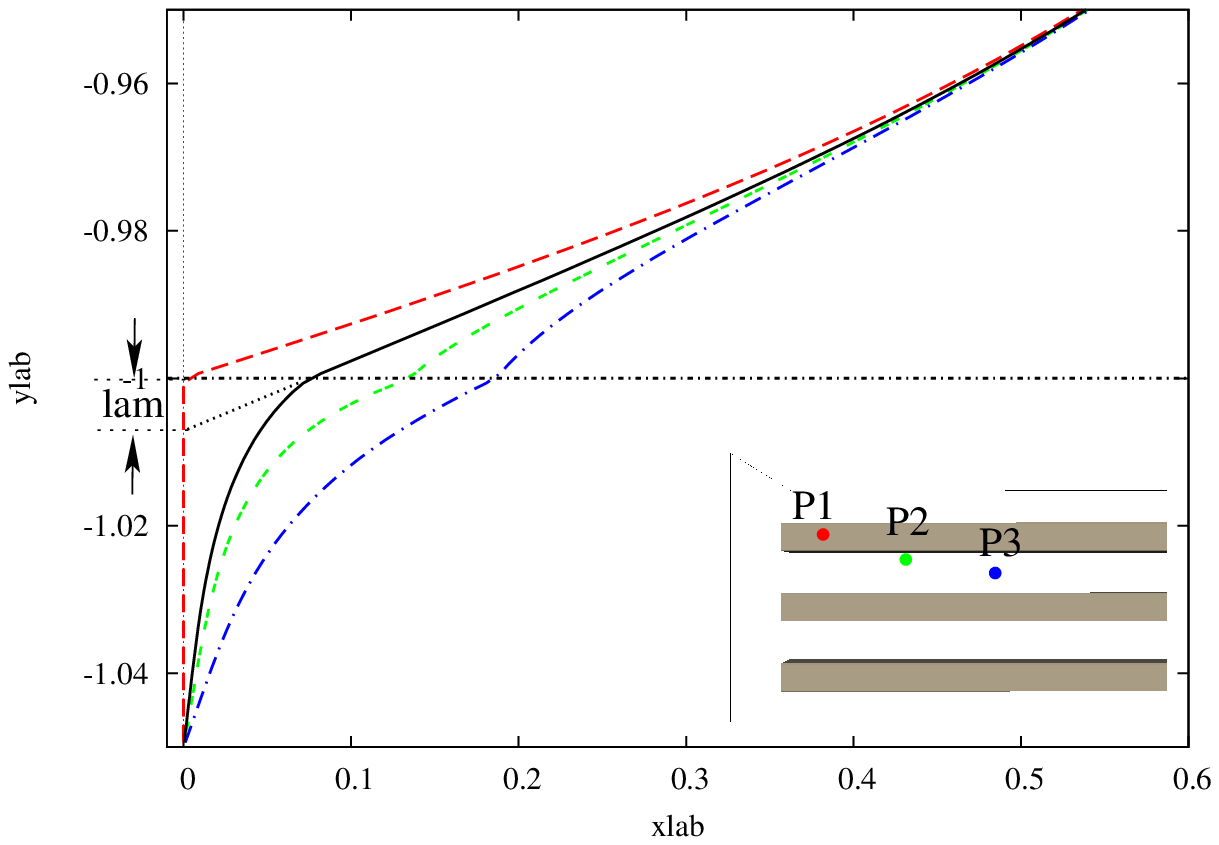}} \\
c)
	\resizebox*{0.45\textwidth}{!}{\includegraphics{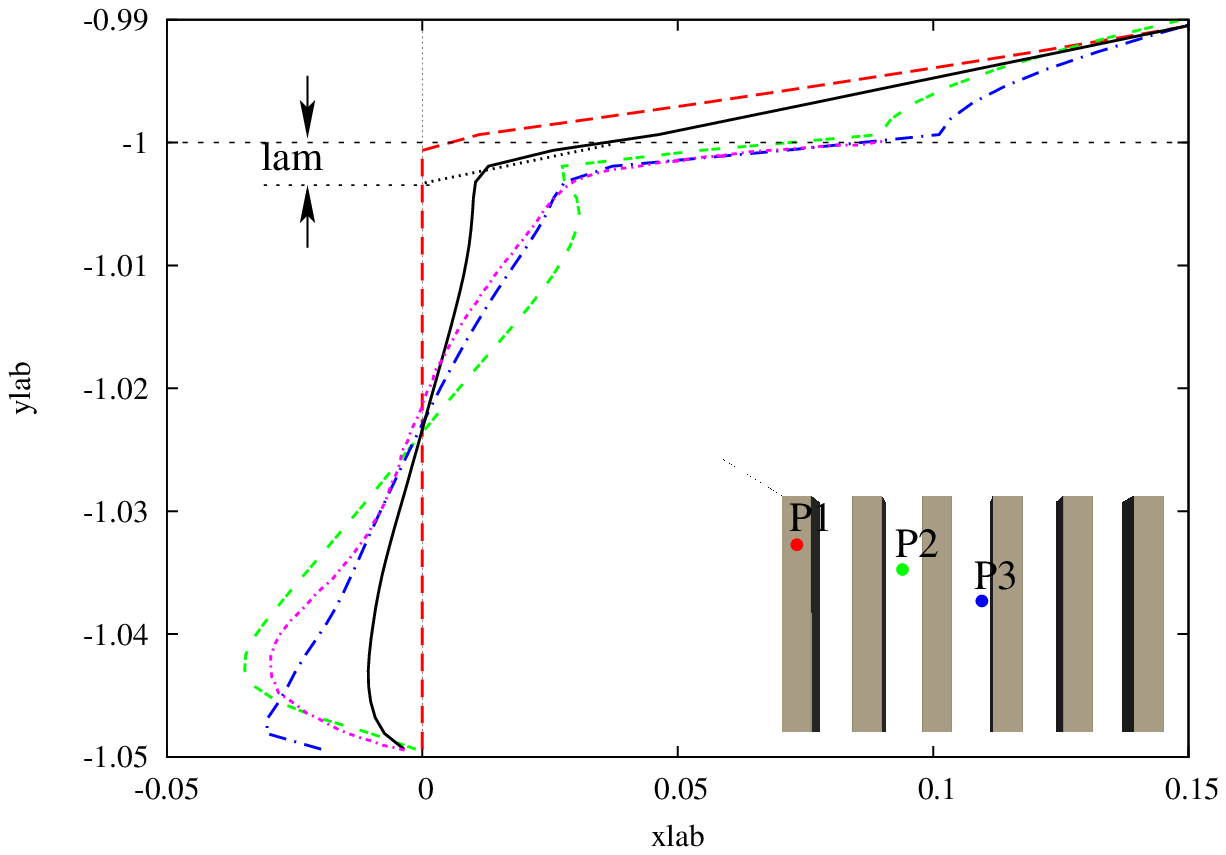}} 
d)
	\resizebox*{0.45\textwidth}{!}{\includegraphics{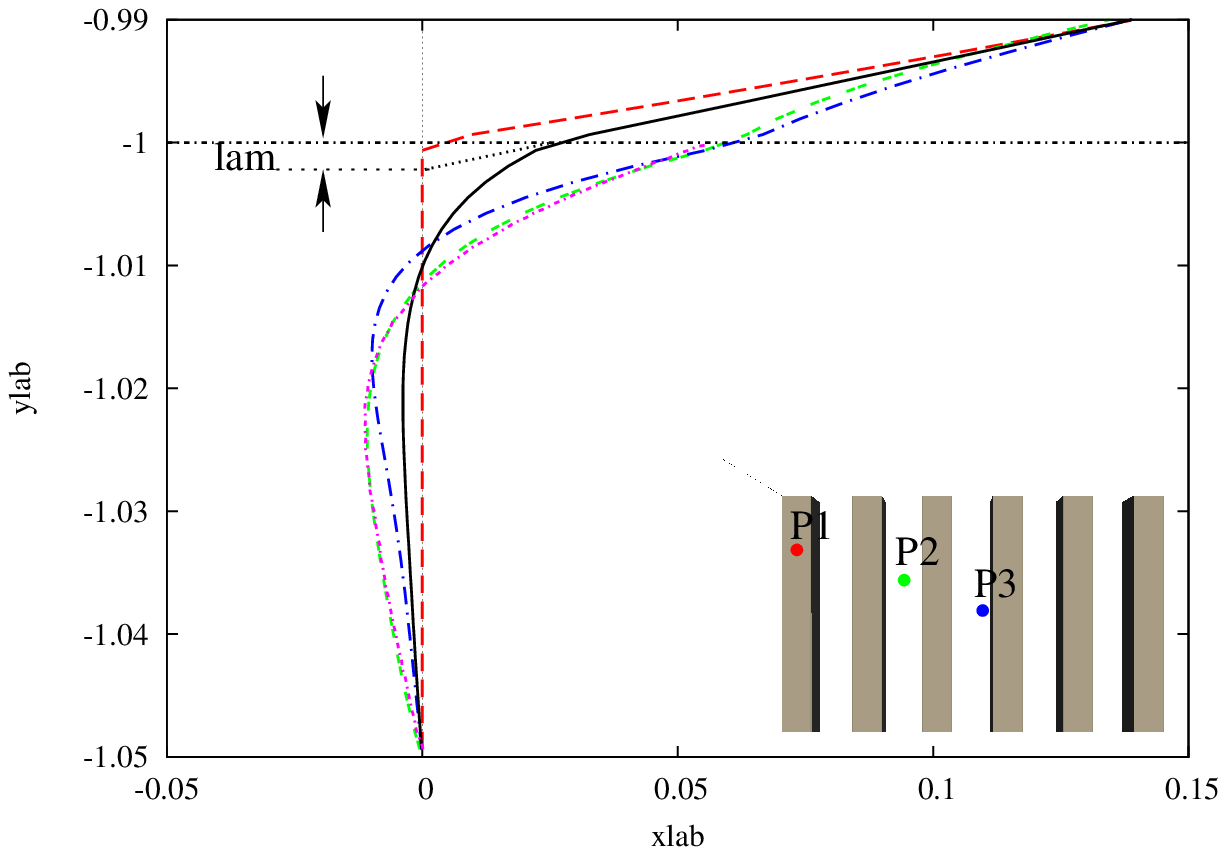}} 
\caption{
	Time averaged streamwise velocity profiles over longitudinal (a,b) and transversal square bars (c,d)
	for  $N=100$ \textbf{a,c)} and  $N=2.5$ \textbf{b,d)}:
	(\textcolor{red}{\dashed})  on the crest ($P_1$);
	(\textcolor{green}{\dashedb}) at a distance $0.0075h$ from the wall  ($P_2$) 
	(the cavity width is $0.05h$);
(\textcolor{blue}{\dashdot}) at the center of the cavity  ($P_3$);
	(\textcolor{black}{\solid}) velocity is also averaged in spanwise and streamwise direction. For transversal square bars,
 velocity profiles at the center of a lid-driven cavity from \cite{Shankar2000} are included as reference (\textcolor{magenta}{\dashdot}).
 }
\label{fig:vel_long}
\end{center}
\end{figure}

Time averaged streamwise velocity profiles ($\langle U \rangle$) at three different positions within the cavities of
longitudinal and transversal square bars are shown in Figures \ref{fig:vel_long} for $N=2.5$ and $100$.
Velocity profiles at the different positions within the cavity converge to the 
time and space averaged velocity profile $\overline{U}(y)$ (black line) for $y/h>-0.96$ for longitudinal grooves and for $y/h>-0.99$ for transversal square bars. This corresponds to about
$k$ and $0.2k$ above the interface ($k$ is the height of the bars).
Because the shear stress at the interface ($y/h=-1$) is continuous, there is a discontinuity in the interfacial velocity gradient, 
with the ratio between the velocity gradient above and below the interface 
inversely proportional to $N$.
The velocity inside the cavities increases with larger values of $N$ (smaller viscosity)
and in the case of longitudinal bars compared to transversal bars. For $N=100$, the velocity profile inside the cavity exhibits
a change of concavity compared to profiles where $N=2.5$.

For transversal bars, the velocity profiles in the center of the cavity 
 compare well with those relative to a lid driven cavity obtained by 
 \cite{Shankar2000} (\textcolor{magenta}{\dashdot} Fig. \ref{fig:vel_long})
 at approximately the
 same cavity Reynolds numbers, $Re_{cav}=U_c k/\nu \simeq 20$ and $1,285$ for $N=2.5$ and $100$ respectively, where $U_c$ is the velocity at the center of the cavity ($P_3$) for $y/h=-1$.

The velocity at the interface averaged in time and space gives the apparent slip 
velocity $U_S=\overline{U}|_{y/h=-1}$ as in \citet{LaugaStone2003}. 
The slip length $\lambda=U_S/(d\overline{U}/dy)|_{i}$ is the distance at which the velocity 
would be zero when extrapolating the interfacial velocity gradient of the overlying fluid into the surface. 
\begin{figure}
\begin{center}
\psfrag{xlab1}{\relsize{+2} $N$}
\psfrag{ylab2}{\relsize{+2} $\lambda/p$, $\lambda/p_c$}
\psfrag{P3}{$P_3$}
a)
\resizebox*{0.45\textwidth}{!}{\includegraphics{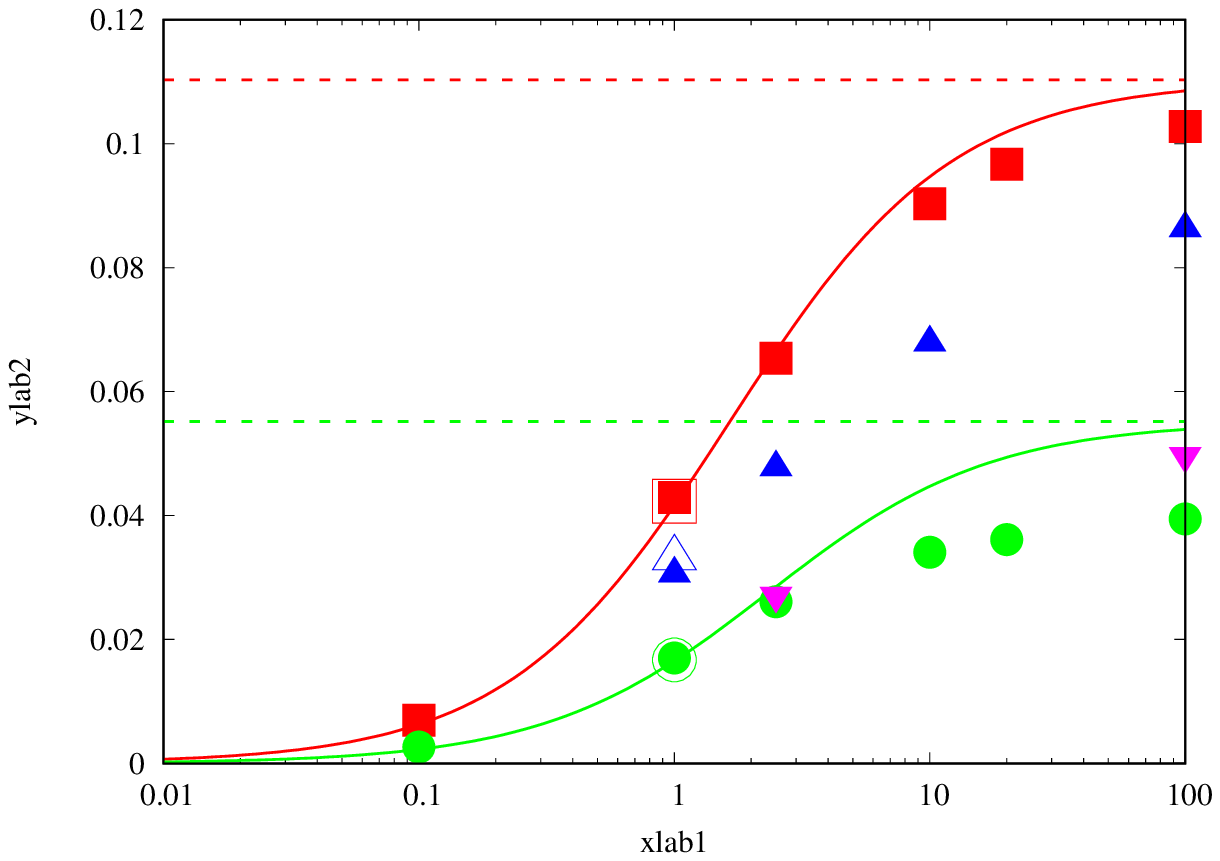}} 
\psfrag{ylab2}{\relsize{+2} $U_s /U_{b}$}
b)
\resizebox*{0.45\textwidth}{!}{\includegraphics{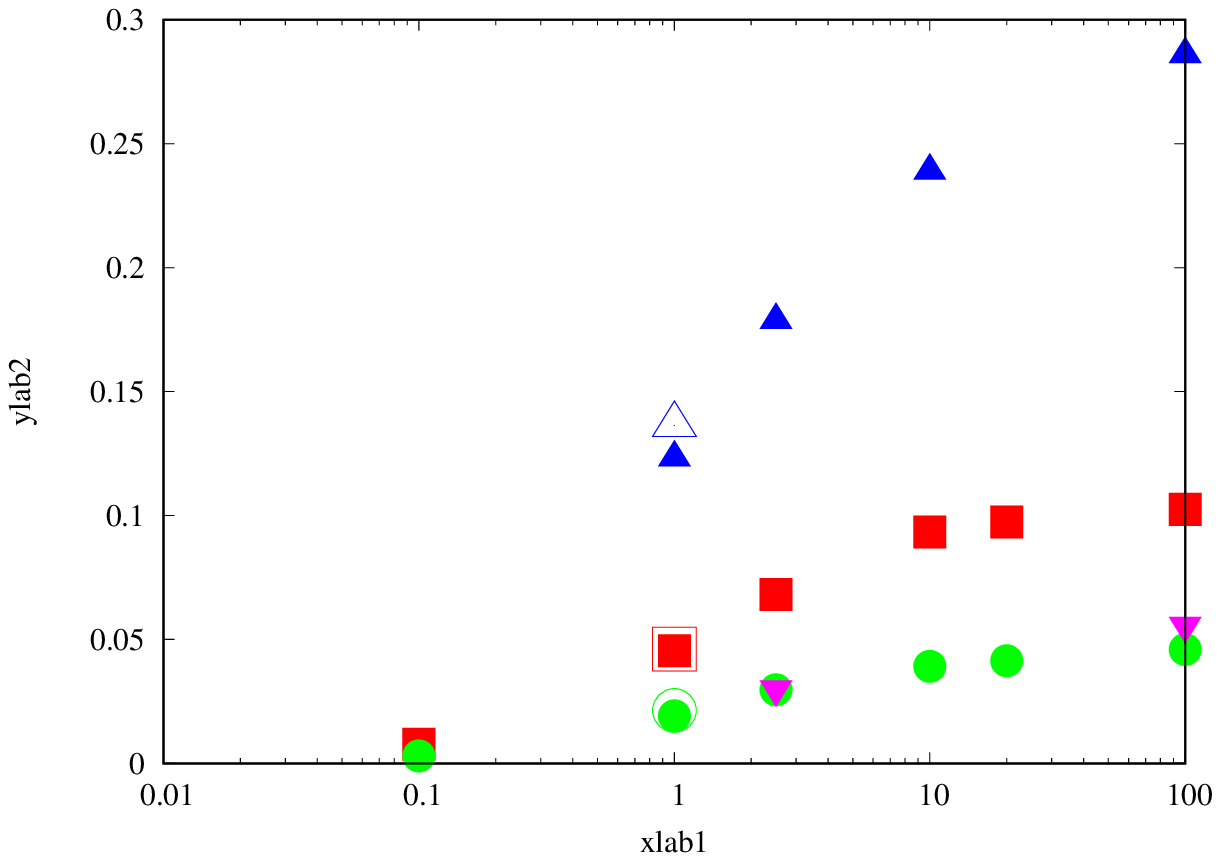}}
\end{center}
	\caption{ Dependence of the slip length (a) and slip velocity (b) on the viscosity ratio.
Lines,
analytical models from \cite{Schonecker2014} (solid) and \cite{Philip1972} (dashed), symbols DNS results:
	{\color{red} \solid}, {\color{red} \dashed}, {\color{red} $\blacksquare$} longitudinal bars; 
	{\color{green}\solid}, {\color{green} \dashed}, {\color{green} $\bullet$} transversal square bars;
	{\color{magenta} $\blacktriangledown$} staggered cubes $a=0.5$, {\color{blue} $\blacktriangle$}  $a=0.875$. Filled symbols are relative to two superposed fluids, empty symbols are for a single phase only.
The slip length relative to arrays of cubes has been normalised by $p_c=k/(1-a)$  i.e. the equivalent pitch of  2D bars  with the same
fluid-area fraction of the cubes.
}
\label{fig:sliplength}
\end{figure}
The apparent slip lengths and slip velocities of the different surface configurations are shown as function of the viscosity ratio in Figure \ref{fig:sliplength}.
The slip velocity and slip length decrease with $N$, i.e. increasing the viscosity of the fluid
in the substrate. Present results 
agree well with the model of 
 \cite{Schonecker2014} for longitudinal bars as discussed in \cite{Fu2017JFM}.
The slip length for the staggered cubes with $a=0.5$ is 
very similar to that of transversal square bars. 
For  larger fluid-area fraction,  $a=0.875$, $\lambda$ becomes much larger than that relative to longitudinal bars. An equivalent pitch can be defined to account for the larger fluid-area fraction of the cubes as
$p_c=k/(1-a)$. This would be the pitch of 2D bars with the same
fluid-area fraction of the cubes.
Scaling the slip length  with the equivalent pitch ($p_c=4p$ for  $a=0.875$), 
$\lambda/p_c$ lies between the longitudinal
and transversal bars, consistently with the geometrical layout which presents longitudinal {alleys} as in longitudinal bars, and obstacles perpendicular to the flow direction as 
in transversal bars.


An additional set of simulations has been performed with the same textured surface as in Fig.1a,b,d but
with
one fluid corresponding to classical longitudinal square bars or rough walls made of staggered cubes or transversal
square bars.
For the flow with a single phase only, the equivalent slip length and slip velocity agree well with those 
obtained with two fluids of same viscosity separated by a slippery interface
($N=1$).
This indicates that the slip length and the reduction of the velocity gradient at the wall are not 
sufficient to explain the reduction of drag, because this same mechanism is present in rough walls too,
which on the contrary increase the drag as it is shown in the next sections.

\begin{figure}
\begin{center}
\psfrag{xlab}{\relsize{+2} \hspace{-1em}$\langle U \rangle/U_b$}
\psfrag{ylab}{\relsize{+2} $y/h$}
	\psfrag{ylab2}{\relsize{+2} ${\langle U_s \rangle}/{U_{b}}$}
\psfrag{lam}{$\lambda$}
\psfrag{xlab1}{\relsize{+2} $z/p$}
a) \resizebox*{0.45\textwidth}{!}{\includegraphics{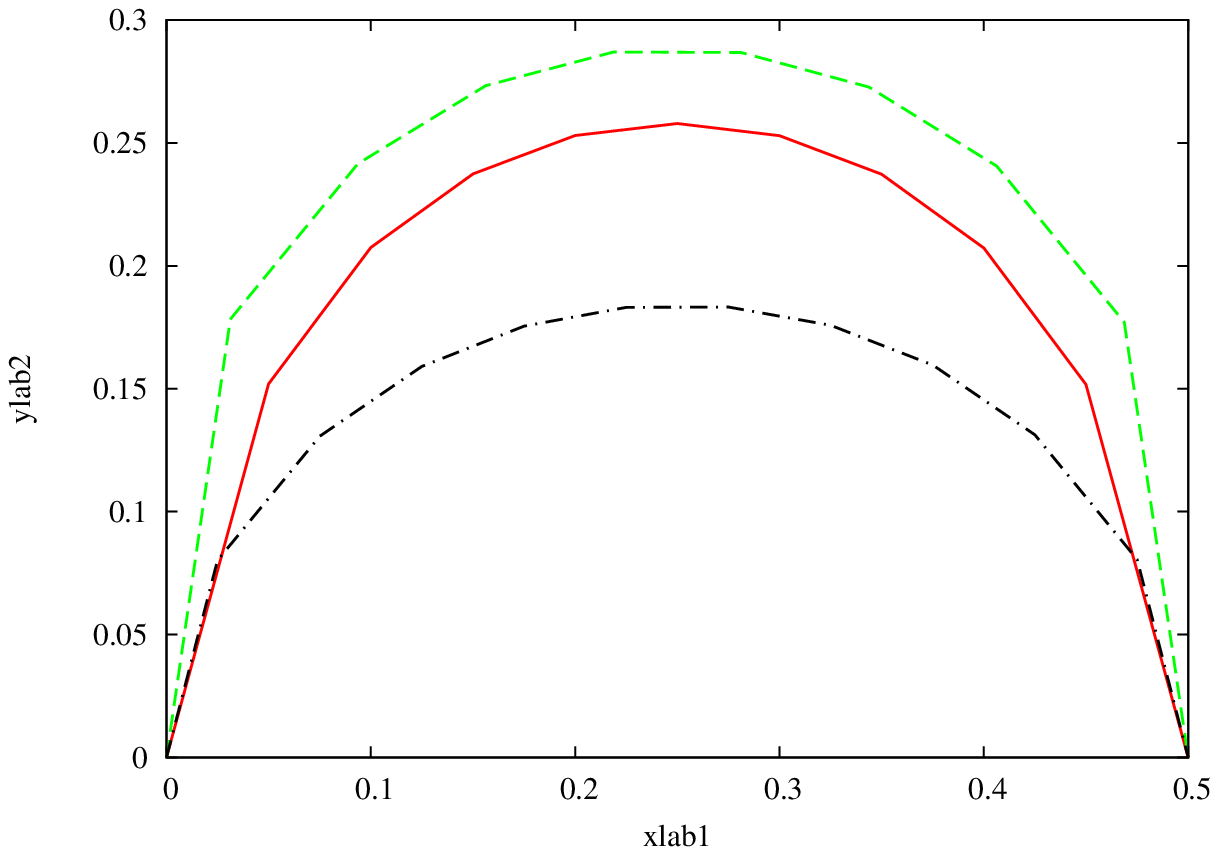}}
\psfrag{xlab1}{\relsize{+2} $x/p$}
b) \resizebox*{0.45\textwidth}{!}{\includegraphics{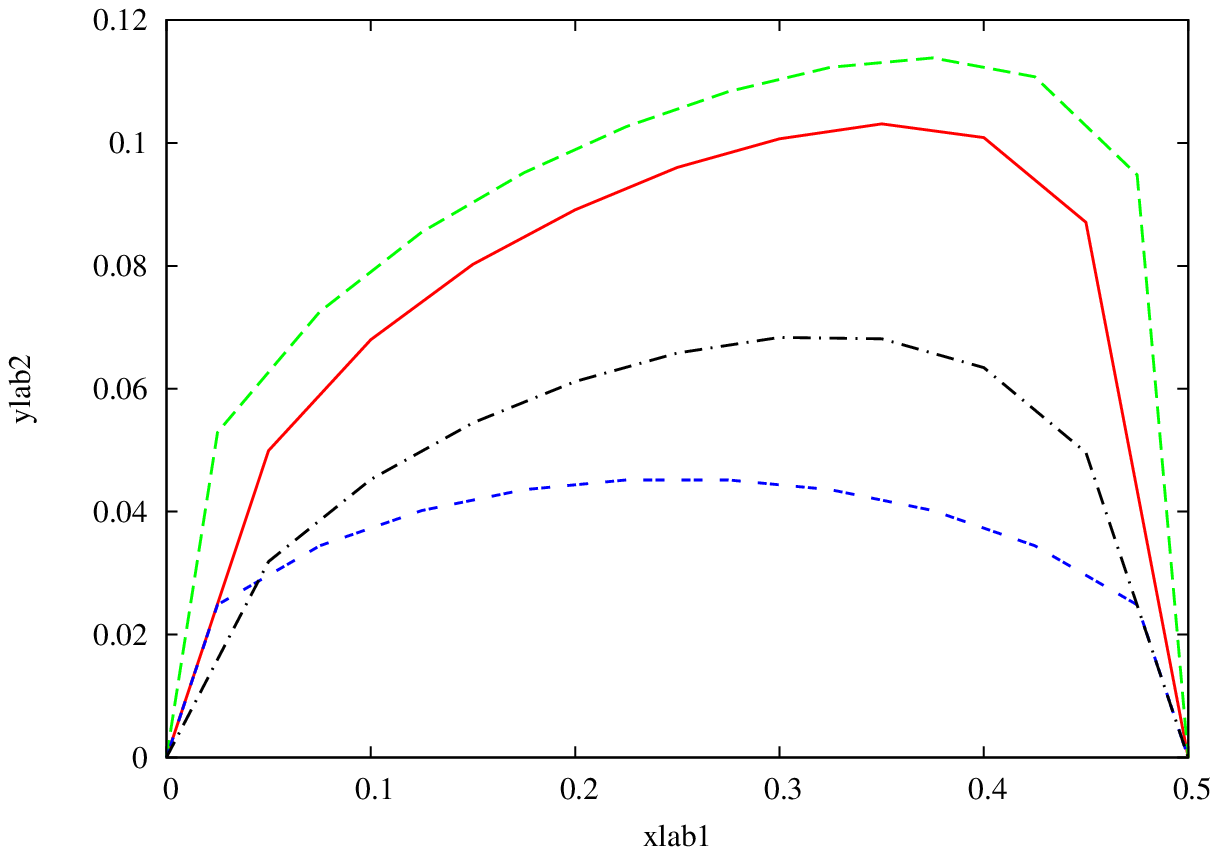}} 
\caption{
	Local slip velocity above longitudinal bars (a) and 
	transversal bars (b): 
	 (\textcolor{green}{\dashed}) free-slip boundary conditions at the interface; 
	  (\textcolor{red}{\solid}) $N=100$; 
	({\color{black} \chndot})   $N=2.5$;
	 (\textcolor{blue}{\dashedb}) Stokes flow.
	}
\label{fig:velocity-crests}
\end{center}
\end{figure}

For longitudinal bars,
the slip lengths for largest values of  $N$ correspond to idealized SHS and 
are in good agreement with the
analytic model of \citet{Philip1972}
and DNS results of \citet{Park2013} (not shown in the figure) which
 modeled SHS as streaks of free-slip and no-slip boundary conditions.
In Fig.\ref{fig:velocity-crests}a, the time averaged velocity at the interface between the two fluids, 
(local slip velocity $\langle U_s \rangle$),
for $N=100$ is compared to that obtained with 
 free slip boundary conditions. The distribution is quite similar, with the local slip velocity
 being slightly smaller in case of $N=100$. 
Since the velocity gradient is smaller too, the resulting slip length is similar to that obtained by
\citet{Park2013}.
Therefore, for these larger viscosity ratios, the contribution of the underlying wall geometry is not 
relevant to determine the slip length and slip velocity.

On the other hand, for transversal square bars, larger differences are observed
at large $N$ with respect to the model by \cite{Schonecker2014}.
This may be because as $N$ increases, the Reynolds number in the cavity increases and the Stokes approximation
is no longer valid.
In Fig. \ref{fig:velocity-crests}b
the velocity at the interface of transversal square bars
obtained solving the Stokes flow equations over the same geometrical setup is included as reference.
While the Stokes' flow velocity distribution on the cavity is symmetric,
the same is not true in general for the turbulent DNS case. The inclusion of
 convective terms at finite Reynolds numbers results in a velocity distribution skewed towards
  the windward edge of the cavity.

\section{Drag Budget }

\begin{figure}
\begin{center}
\psfrag{xlab1}{\relsize{+2} $N$}
\psfrag{ylab2}{\relsize{+2} $(C_{f,c},C_{f,b},P_d)/\tau$}
\psfrag{c1}{\relsize{+2} $A$}
\psfrag{c2}{\relsize{+2} $B$}
\psfrag{c3}{\relsize{+2} $C$}
\psfrag{c4}{\relsize{+2} $D$}
\psfrag{c5}{\relsize{+2} $E$}
\psfrag{c6}{\relsize{+2} $F$}
\psfrag{c7}{\relsize{+2} $G$}
\psfrag{c8}{\relsize{+2} $H$}
\psfrag{c9}{\relsize{+2} $I$}
\psfrag{c10}{\relsize{+2} $L$}
\psfrag{c11}{\relsize{+2} $M$}
\psfrag{c12}{\relsize{+2} $N$}
\psfrag{f1}{\relsize{+2} flow dir.}
\textbf{a)}\resizebox*{0.5\textwidth}{!}{\includegraphics{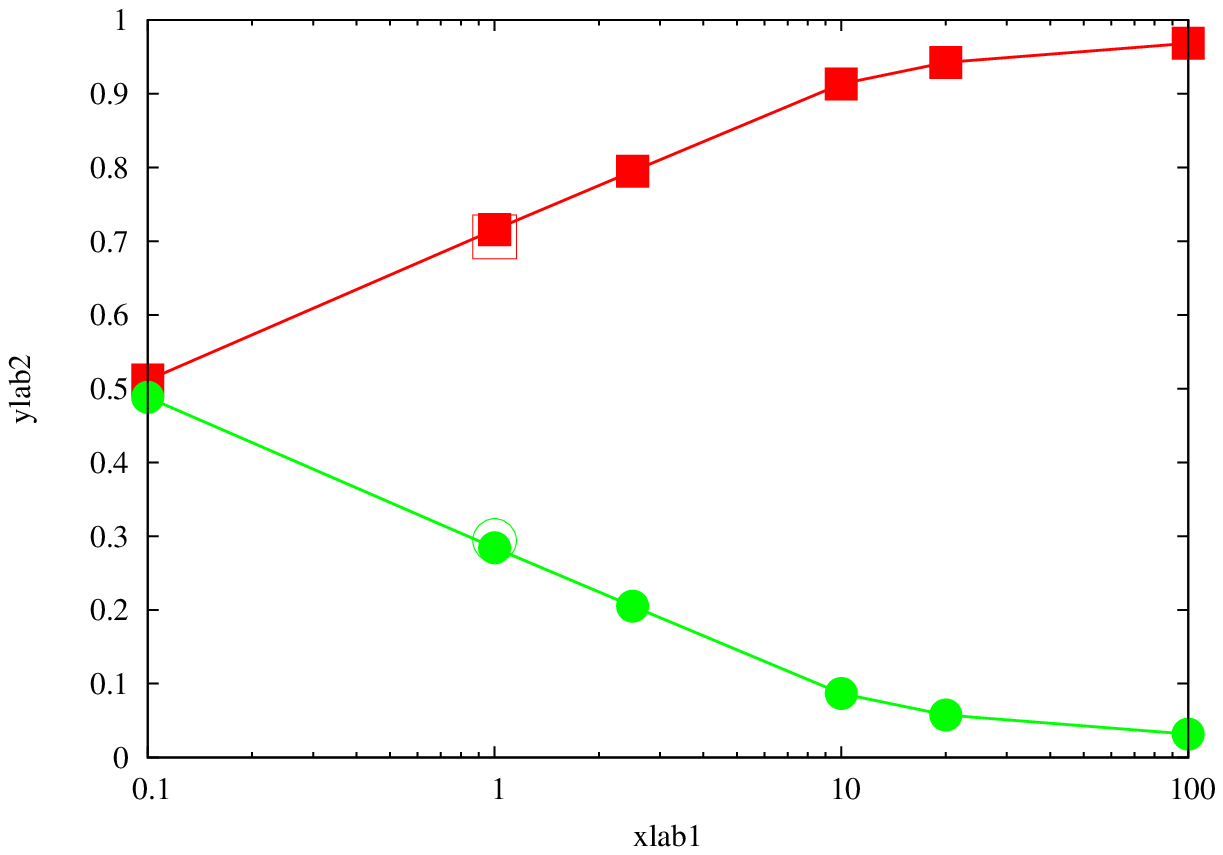}}
\resizebox*{0.4\textwidth}{!}{\includegraphics{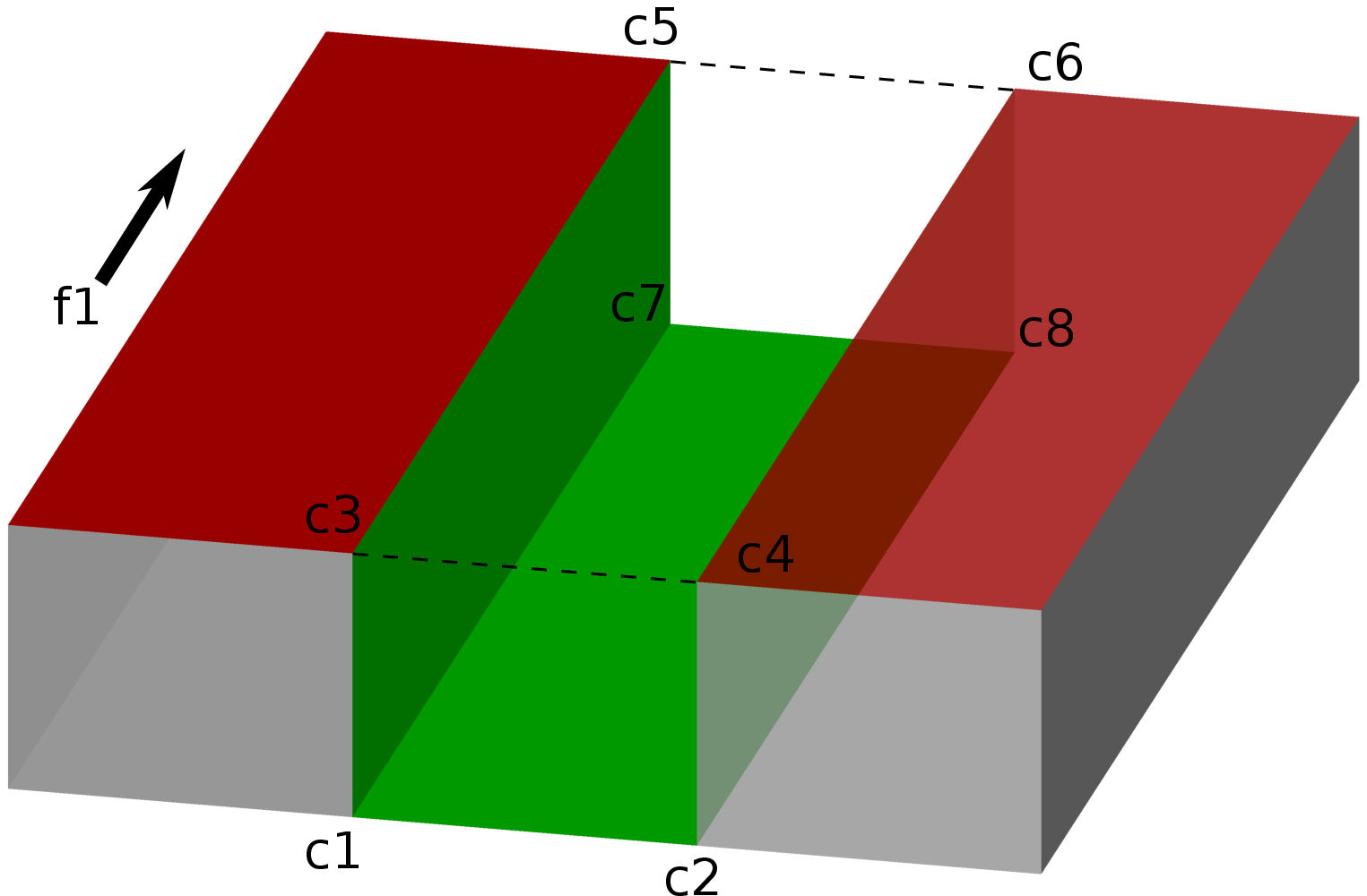}}\\
\textbf{b)}\resizebox*{0.5\textwidth}{!}{\includegraphics{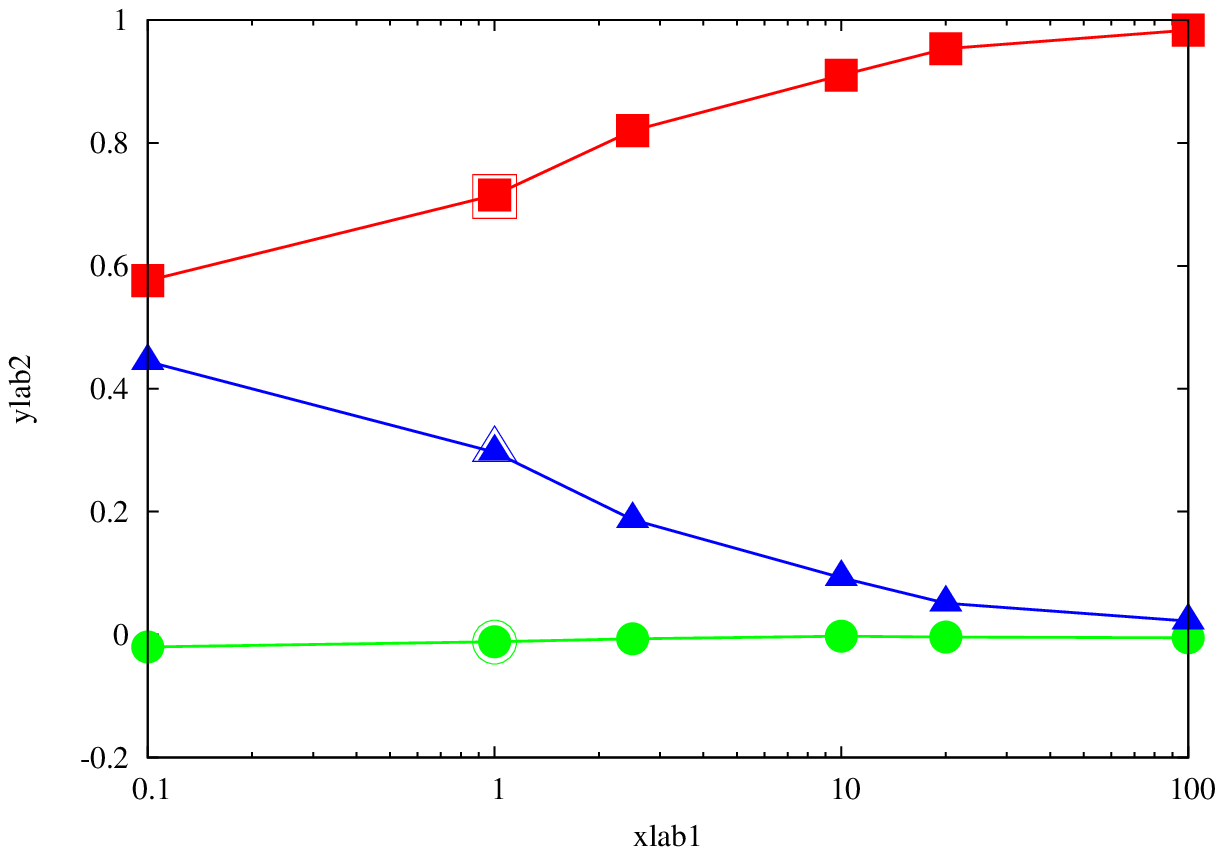}}
\resizebox*{0.4\textwidth}{!}{\includegraphics{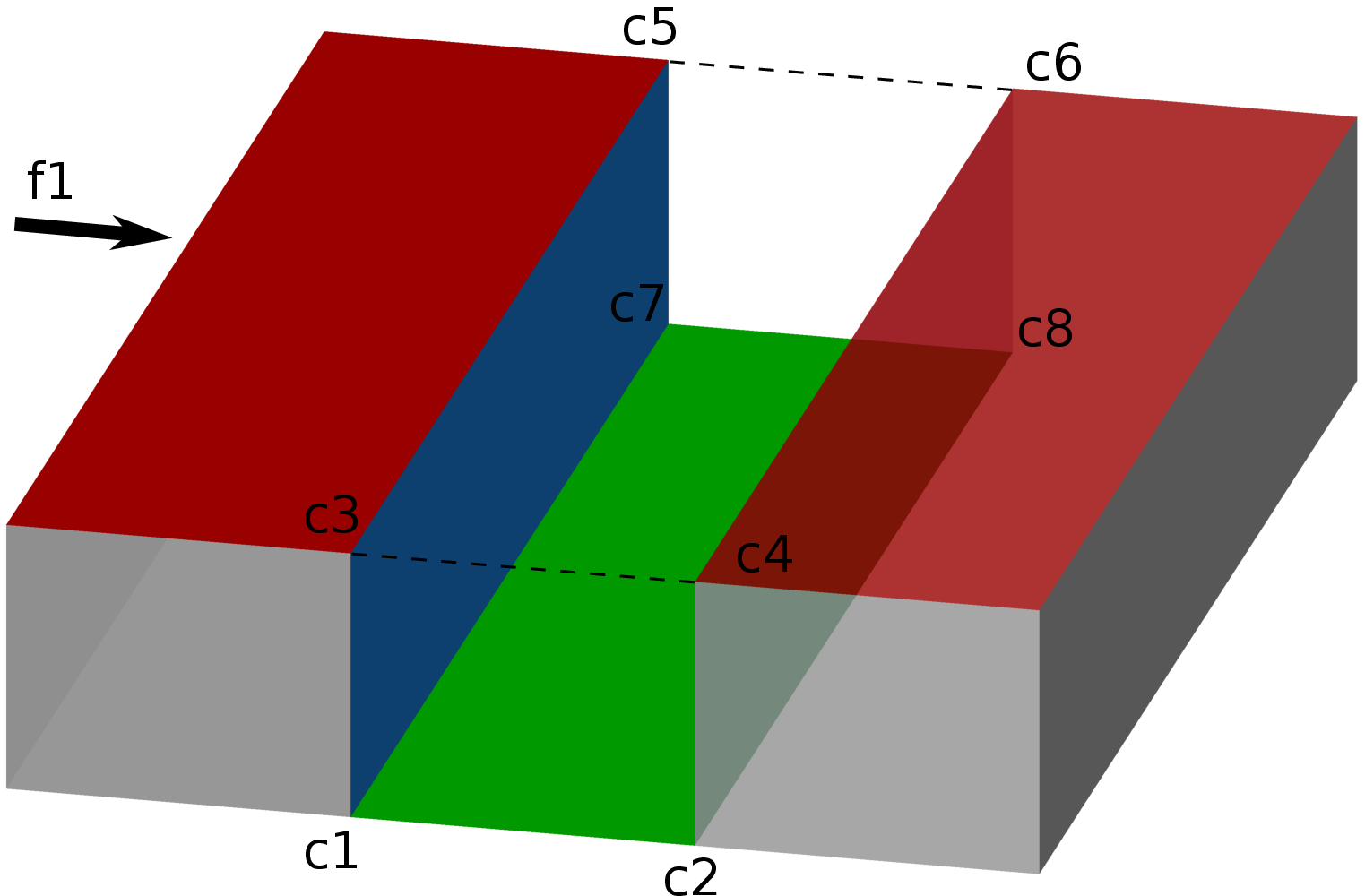}}\\
\textbf{c)}\resizebox*{0.5\textwidth}{!}{\includegraphics{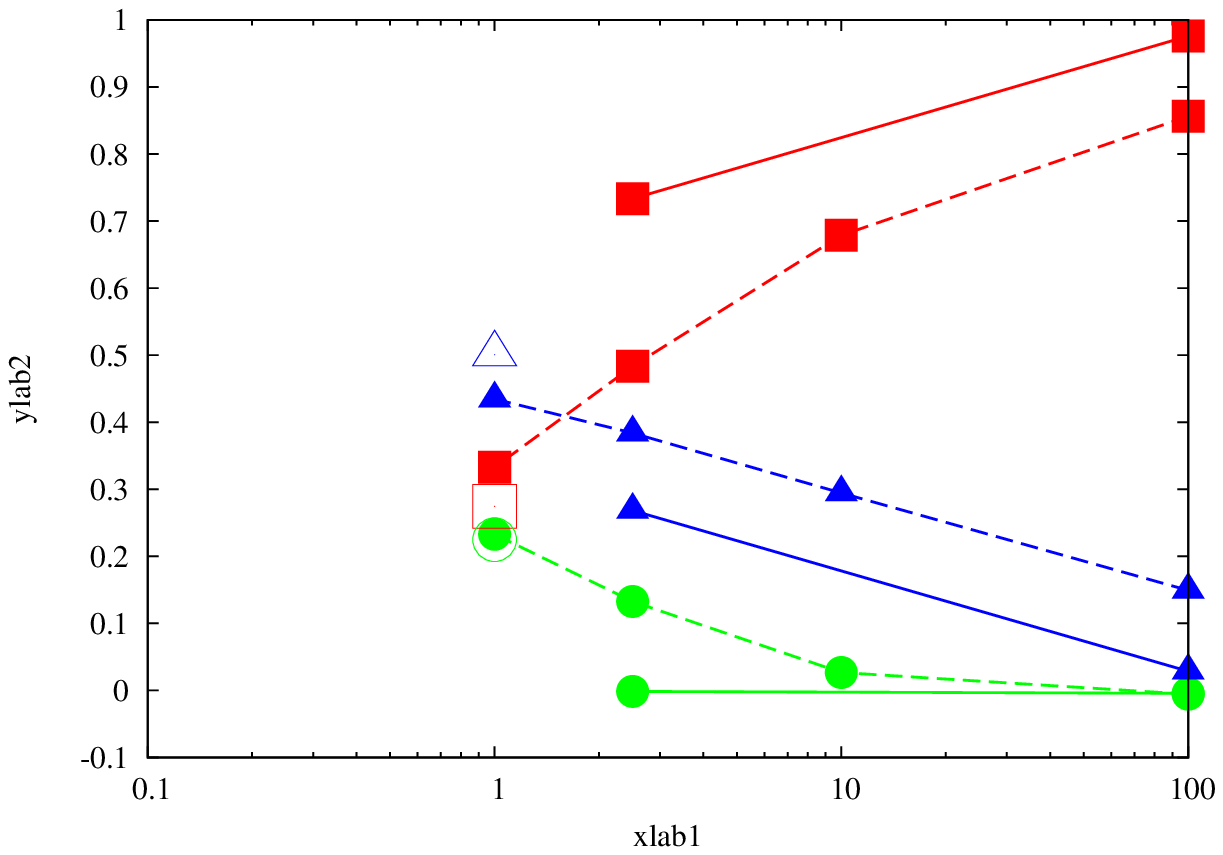}}
\resizebox*{0.4\textwidth}{!}{\includegraphics{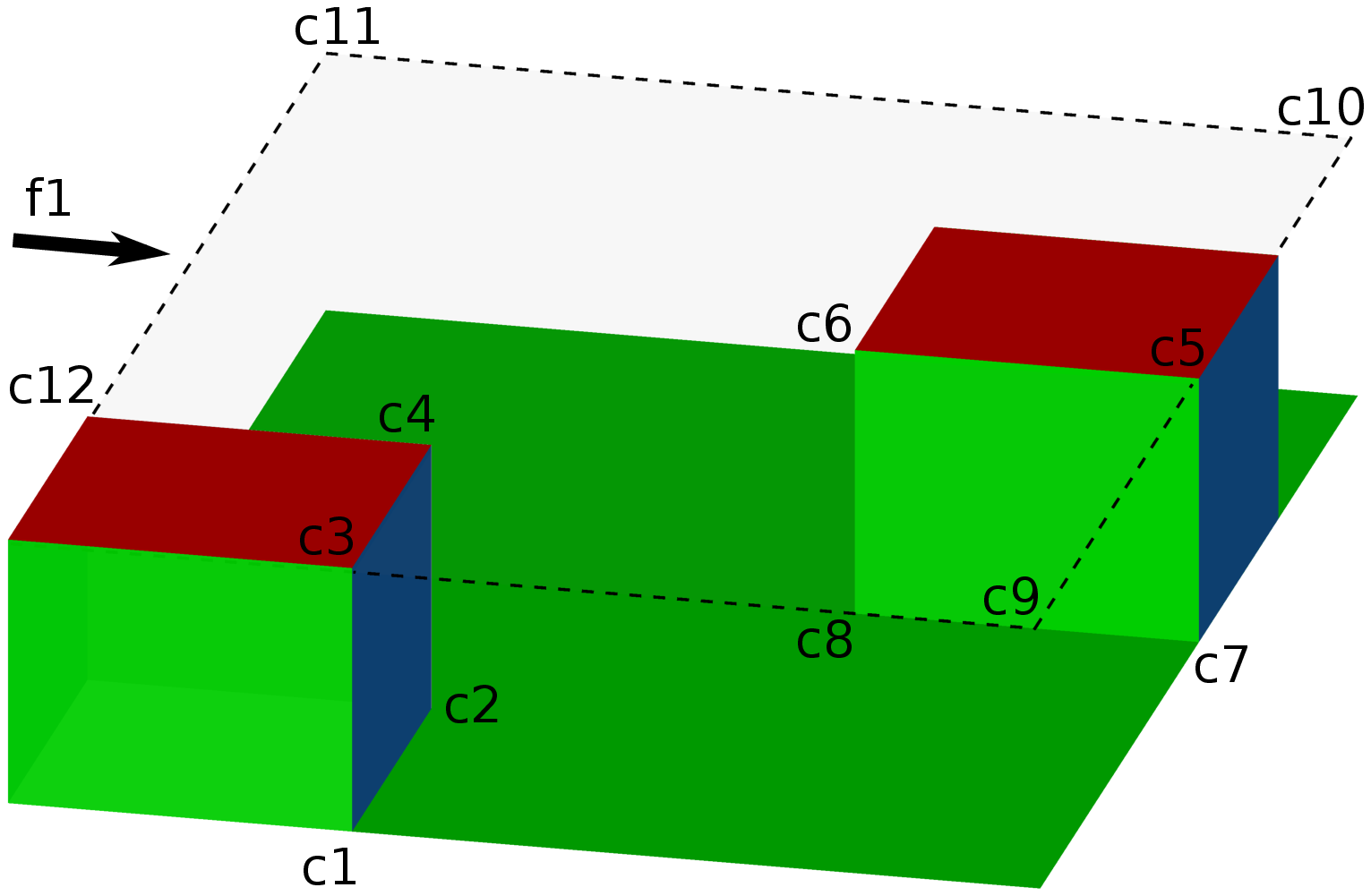}}\\
\end{center}
	\caption{Form drag ({\color{blue} \solid, $\blacktriangle$}),
	frictional drag on the crests plane ({\color{red} \solid}, {\color{red} $\blacksquare$}) and
	inside the cavities ({\color{green} \solid}, {\color{green} $\bullet$}) 
	normalized by the total drag as a function of the viscosity ratio $N$:
\textbf{a)} Longitudinal square bars, \textbf{b)} Transversal square bar and \textbf{c)} Staggered cubes.
Solid lines refer to $a=0.5$, dashed to $a=0.875$. Empty symbols indicate
a single fluid (plotted in correspondence of $N=1$ for analogy), 
	solid symbols two fluids ($N$ varies).}
\label{fig:cf_tau}
\end{figure}
{ 
By integrating the time averaged Navier-Stokes equations for $U$ ($i=1$ in Eq.3.1)  over the fluid volume in the channel 
it is obtained:
\begin{equation}
\int_{V} \left( \frac{\partial \langle U \rangle}{\partial t}
+\frac{\partial{\langle U U_{j} \rangle}}{\partial{x_{j}}} \right) dv
=\int_{V} \left( -\frac{\partial \langle P \rangle}{\partial x}+
\frac{1}{Re} \frac{\partial^{2} \langle U \rangle }{\partial{x_{j}^{2}}} + \langle \Pi \rangle
\right)
dv \; ,
\label{eq:force0}
\end{equation}
where $V$ is the volume of fluid in the channel including the volume of the fluid in the texture.
Since the flow rate is constant ($\int_{V} \partial{\langle U \rangle }/\partial{t}=0$), 
 for  periodicity in $x$ and $z$ and no-slip condition at the walls, Eq. \ref{eq:force0}
 reduces to
\begin{equation}
	\begin{aligned}
		\underbrace{\langle \Pi \rangle  V}_\text{Forcing} =  &
	\underbrace{\int_{S_b} \langle P \rangle \vec{n} \cdot \vec{x} ds}_\text{Pressure drag} -\underbrace{\int_{S}  \frac{1}{Re} \frac{d \langle U \rangle}{d y}|_{y_{up}} ds}_\text{Friction upper wall} + \\
		&	 +\underbrace{\int_{S_{cr}}  \frac{1}{Re} \frac{d \langle U \rangle}{d y}|_{y_{cr}} ds}_\text{Friction crests} +\underbrace{\int_{S_{cav}}  \frac{1}{Re_1} \frac{d \langle U \rangle}{d n}|_{cav}  \vec{x} \cdot \vec{ds}}_\text{Friction cavity} \; ,
	\end{aligned}
\label{eq:forcing}
\end{equation}
where $y_{up}$ corresponds to the upper wall, $y_{cr}$ is the crests plane,
$S_{cr}$ is the solid surface at the crests plane of the  texture (red in Fig.5), 
$S$ is the area of the upper wall (smooth),
 $S_{cav}$ indicates the walls of the cavities aligned to the flow (green in Fig.5), 
 $S_b$ the walls of the cavity perpendicular to the flow direction (blue in Fig.5),
 $\vec{n}$ is the normal to the walls in the cavity, $\vec{x}$ is the unit vector  in the $x$ direction, 
 $\vec{ds}$ is the vector tangential to the walls of the cavity and
$Re_1=N Re$ is the Reynolds number of the fluid in the cavities.
Dividing Eq.\ref{eq:forcing} by $S$, we obtain that the forcing required to keep the flow rate constant ($\langle \Pi \rangle  V/S$) is equal to the sum of the 
form drag of the texture ($P_d$), 
friction on the upper wall, $C_{f,up}$, friction on the crests plane $C_{f,c}$ and friction on the bottom wall $C_{f,b}$.
\begin{align}
	{P_{d}}& =S^{-1} \int_{S_b} \langle P \rangle \vec{n} \cdot \vec{x} ds  \; ,\\
	C_{f,up} &=-{S^{-1}} {\int_{S}  \frac{1}{Re} \frac{d \langle U \rangle}{d y}|_{y_{up}} ds}  \; ,\\
        C_{f,c}&= {S^{-1}}  \int_{S_{cr}}  \frac{1}{Re} \frac{d \langle U \rangle}{d y}|_{y_{cr}}\label{eq:drc} ds \; ,\\
	C_{f,b}&= {S^{-1}} \int_{S_{cav}}  \frac{1}{Re_1} \frac{d \langle U \rangle}{d n}|_{cav}  \vec{x} \cdot \vec{ds} \label{eq:drb} \; ,
\end{align}
(note all  the quantities are non dimensional since derived from Eq.3.1, therefore lengths are normalised 
by $h$, velocities by $U_b$ and pressure by $\rho U_b^2$).

}

The friction on the crests plane, in the cavities and
the form drag, normalized by the total drag are shown in Figure \ref{fig:cf_tau}. 
For $N>50$, the contribution of the friction below the crests plane and the form drag
is,  to a good approximation, negligible for transversal square bars. For the case of longitudinal square bars, the
form drag is identically zero since there are no walls perpendicular to the flow direction.
The friction on the side walls and bottom walls is about $3-4\%$ of the total drag.
This may indicate that the assumption of free slip boundary conditions on the crests plane may overestimate the 
amount of drag reduction of such quantity.

While previous studies have focused mostly on $N=\infty$, here we extend the analysis to lower values of $N$ which
represents lubricants with higher viscosity. 
As $N$ decreases, the value of the friction and form drag in the texture
increases. 
This is because there is a larger
momentum transfer inside the cavities balanced by 
an increase of frictional and pressure drag. 

For large values of $N$, $Re^{-1} {d \langle \overline{U} \rangle}/{d y}$
and then ${P_{d}} +{C_{f,b}}$ are very small. 
Reducing $N$ decreases the slip velocity (see Figure \ref{fig:sliplength}) and increases
the velocity gradient and momentum transfer at the interface. Consequently,
 the form drag and friction inside the substrate increase as well. 
The relative contributions to the total drag of ${P_{d}}$ and 
${C_{f,b}}$ are highly dependent on the particular layout of the substrate.
For example,
in the case of transversal bars, the value of the friction inside the cavities remains close to zero for all values of
$N$. This is because of the recirculation and reverse flow on the bottom of the cavity.
On the other hand, for longitudinal bars, a decrease of $N$ corresponds to a significant increase of the friction below the crests plane. For very small values of $N$, $C_{f,c}$ and $C_{f,b}$ are almost equal.
Staggered cubes with $a=0.5$ present a drag breakdown similar to transversal bars. The recirculation between two cubes
is quite weak as between two consecutive transversal bars and may explain the similarities between the two surfaces
in terms of slip length, slip velocity, and drag breakdown. 
The cubes with gas fraction $a=0.875$, on the other hand, have a larger
form drag, due to the larger momentum transferred  inside the substrate because
of the larger area of the interface compared to $a=0.5$. 
Even for $N=100$, the contribution of the form drag is about $15\%$ of the 
total drag, therefore, modeling the surface with free slip boundary conditions,
neglecting the drag in the substrate does not seem to be appropriate for this
layout.

The values of $C_{f,c},C_{f,b},P_d$  relative to one fluid only
are also shown in Figure  \ref{fig:cf_tau} (empty symbols) in correspondence to $N=1$ (since there is only one value of viscosity inside and outside the 
substrate). 
 The drag breakdown is approximately the same as that of two fluids with same viscosity and a slippery interface
(no wall normal velocity, only slip velocity) for both transversal and longitudinal bars and only slightly different for staggered cubes.
In addition, the dependence of $C_{f,c},C_{f,b},P_d$ with $N$ is continuous, implying that
changes to the amount of 
drag reduction do not reflect a  different mechanism, but are due to
the amount of momentum transferred from
the bulk flow to  the cavity.

\begin{figure}
\begin{center}
\psfrag{xlab1}{$N$}
\psfrag{ylab2}{$1-\tau/\tau_0$}
\includegraphics[width=0.7\textwidth]{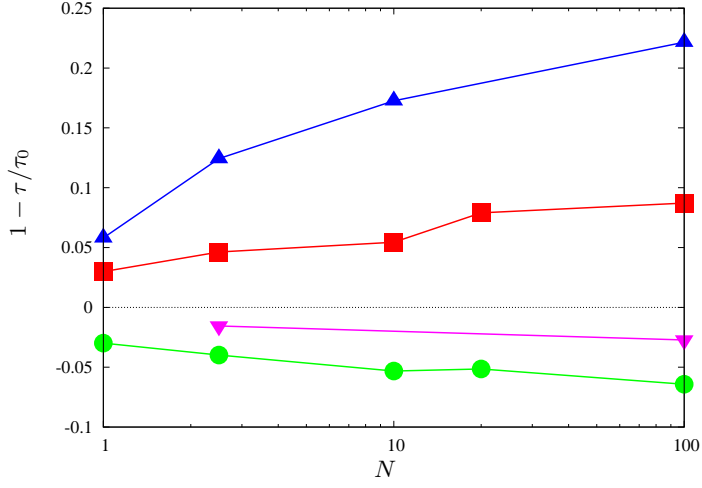}
\end{center}
\caption{Dependence of the drag reduction  on the viscosity ratio ($N$):
{\color{red} $\blacksquare$} longitudinal bars, {\color{green}  $\bullet$} transversal square bars, staggered cubes,
{\color{magenta}  $\blacktriangledown$ } $a=0.5$, {\color{blue} $\blacktriangle$}  $a=0.875$.
}
\label{fig:N_dr}
\end{figure}

To quantify the amount of drag reduction, simulations of a smooth channel at the same Reynolds number (based on the viscosity of the
main stream) and same flow rate were carried out.
The amount of drag reduction  $DR=(\tau_0-\tau)/\tau_0$ ($\tau_0$ indicates the drag of the smooth channel)
as a function of $N$,
is shown in Fig. \ref{fig:N_dr}, a negative value implying drag increase.
The value of $\tau$, calculated as  $\tau=C_{f,c}+C_{f,b}+{P_d}$
 is further validated by subtracting
the friction on the upper
smooth wall
to the total pressure drop of the channel
and by extrapolating the total shear stress 
($\overline{uv}+\mu d\overline{U}/dy$) at the wall ($\overline{uv}$ is the Reynolds stress as defined in Eq. \ref{eq:Restress2} including both the dispersive and incoherent component). 
The differences in the 3 values of $\tau$ obtained are smaller than $0.5\%$. 
Because our parameterization covers several orders of magnitude in N, this small uncertainty in $\tau$ does not affect the  general conclusions.
For a fluid--area fraction $a=0.5$, only longitudinal square bars were found to reduce the drag with respect to a smooth channel. 
Both staggered cubes and transversal bars increase the drag by $2 \%$ and $5\%$ respectively. 
The drag for these two geometries does not depend significantly on the viscosity ratio. As $N$ increases, the friction and form drag (for the cubes) in the cavities decrease. This, however, is offset by an increase in 
friction on the crests plane as a consequence of the increased slip velocity.
On the other hand, the drag reduction increases significantly by increasing $N$ for 
both longitudinal bars and staggered cubes 
with $a=0.875$, the latter being the most effective in terms of reducing the drag given the larger 
fluid-area fraction.
For these two geometries, the drag reduces with $N$ because the reduction of
the friction on the bottom wall 
dominates and each can support a substantial streamwise flow within the  lubricating layer.

It is perhaps surprising that drag reduction can be obtained not only with a significantly less viscous superposed fluid (i.e. $N \gg 1$) inside the textured surface, but even with $N\simeq 1$. 
{ For example, for $N\simeq 2.5$, corresponding to the viscosity ratio of water over heptane,
the drag over a substrate made of longitudinal bars or staggered cubes
is reduced by about $5 \%$ and $12\%$ respectively.
}
Even for $N=1$, which indicates two fluids with the same viscosity separated by a slippery flat interface, the drag is reduced.
This is perhaps counter intuitive because the flow configuration is very similar to a  rough surface
made of staggered cubes or  longitudinal square bars which normally increase 
the drag, (see for example \cite{Leonardi2010}).
In fact, we performed DNS over
the same texture with a single fluid without interface and found a higher drag
with respect to that obtained with two fluids separated by an interface and $N=1$ (Table 1).
For longitudinal
  square bars  the drag is approximately the same as that of a smooth wall,
  while for
 transversal bars and 
staggered cubes the drag increases by $18 \%$ and $64\%$ respectively.
\begin{table}
  \begin{center}
  \begin{tabular}{l|c|c} \hline
  & two fluids $N=1$ with interface  & one fluid only\\\hline\hline
 Longitudinal square bars&$0.03$ & $0$\\
 Transversal square bars&$-0.03$ & $-0.18$\\
 Staggered Cubes $a=0.875$& $0.06$ & $-0.64$\\\hline
  \end{tabular}
  \end{center}
  \caption{Comparison of the amount of drag reduction obtained by two superposed fluids with the same viscosity and a slippery interface ($N=1$) with that of a single fluid over the same textured surface.}
  \label{t:unitsym}
\end{table}

{ 
\begin{figure}
\begin{center}
\psfrag{xlab}{\relsize{+2} $U_s/U_b$}
\psfrag{DR}{\relsize{+2} DR}
a)
\resizebox*{0.45\textwidth}{!}{\includegraphics{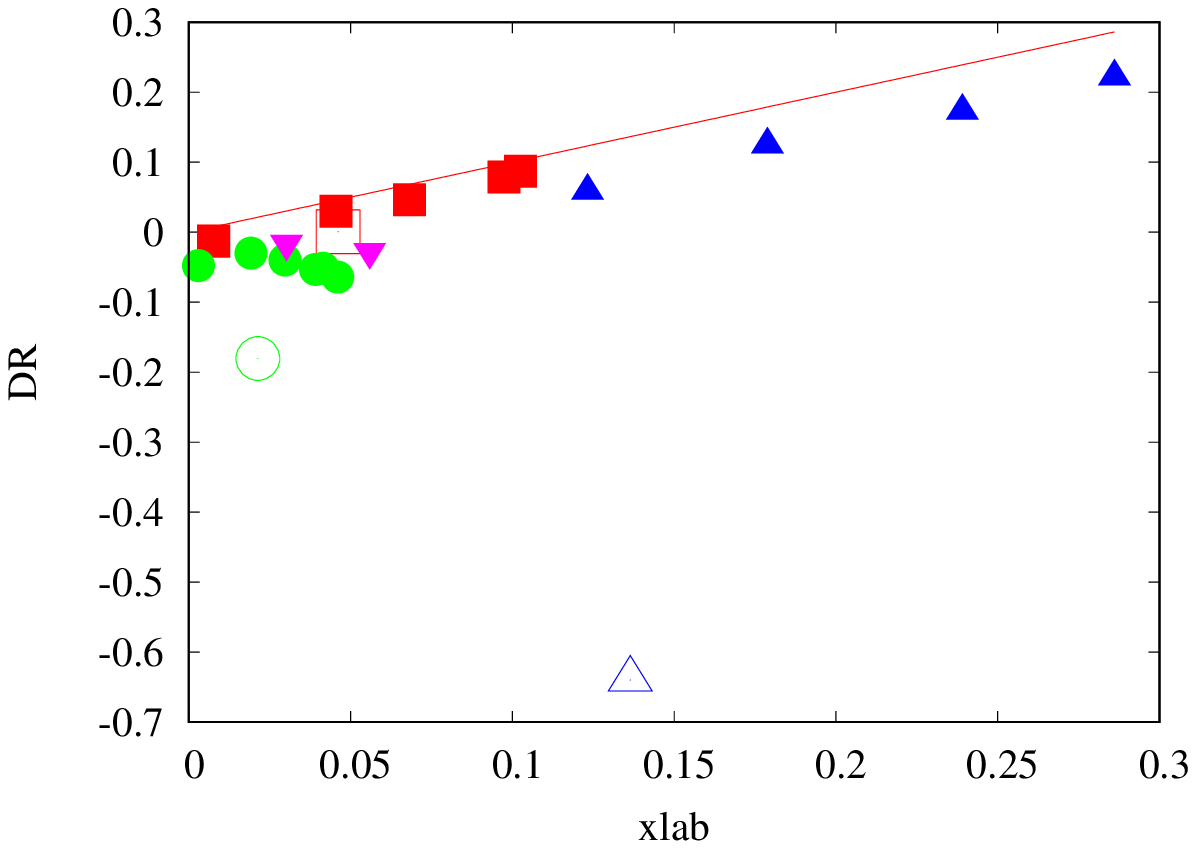}} 
b)
\psfrag{xlab}{\relsize{+2} $\lambda^+$}
\resizebox*{0.45\textwidth}{!}{\includegraphics{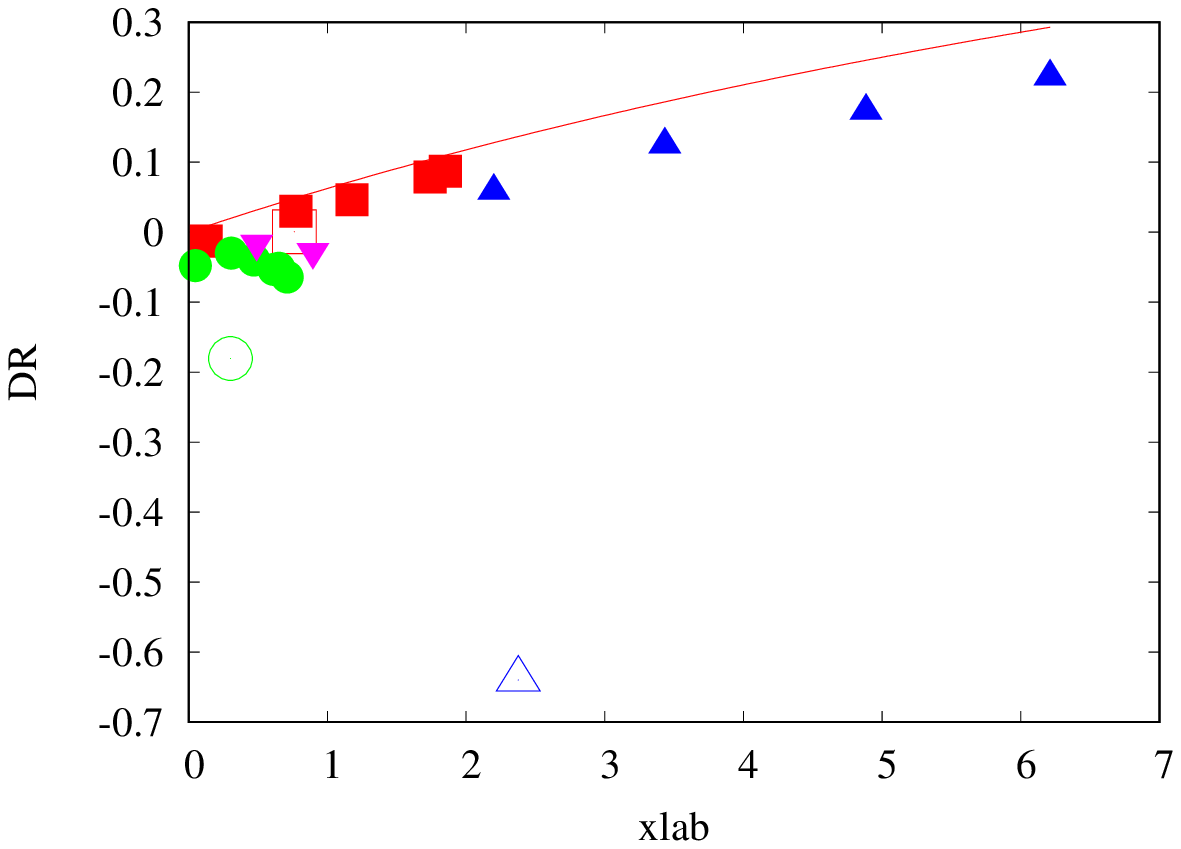}}
\end{center}
	\caption{ Drag reduction as function  of the slip velocity (a) and slip length in wall units (b):
	 {\color{red} $\blacksquare$} longitudinal bars; 
	 {\color{green} $\bullet$} transversal square bars;
	{\color{magenta} $\blacktriangledown$} staggered cubes $a=0.5$, {\color{blue} $\blacktriangle$}  $a=0.875$. Filled symbols are relative to two superposed fluids, empty symbols are for a single phase only. Solid lines are equations \ref{eq:rastegari1} and \ref{eq:rastegari2} respectively.
}
\label{fig:DR-Us-Lam}
\end{figure}
This inconsistency between the results with $N=1$ and a single fluid is also observed
 in the dependence of the drag on the slip velocity and  slip length.
\cite{Rastegari2015} derived an analytical expression correlating the amount of 
drag reduction with either the slip velocity or the slip length:
\begin{equation}
	DR=\frac{U_s}{U_b}+O(\epsilon) \;,
	\label{eq:rastegari1}
\end{equation}
\begin{equation}
	DR=\frac{\lambda^+}{\lambda^+ +Re/Re_\tau }+O(\epsilon) \;.
	\label{eq:rastegari2}
\end{equation}
where $U_s/U_b$ and $\frac{\lambda^+}{\lambda^+ +Re/Re_\tau }$ represents the contribution of either 
the slip velocity or length  to the drag reduction and the term 
$O(\epsilon)$ the modification of the turbulence structure and secondary flow.
Present results for both SHS, LIS and rough surfaces (one fluid only)
are shown in Fig. \ref{fig:DR-Us-Lam}. For both idealised SHS and LIS,
numerical results agree well with Eqs.\ref{eq:rastegari1} and 
\ref{eq:rastegari2} when the texture is made of longitudinal bars and 
staggered cubes. For the latter,  data are shifted about  $5\%$
below the analytical correlation
but the trend is very similar. 
On the other hand, the drag obtained for transversal bars  does not correlate well with slip velocity and slip length because the term $O(\epsilon)$ overcomes the modest reduction of drag due 
to the reduced shear on the cavity.
In case of a single fluid, the amount of drag reduction for cubes and transversal bars is 
inversely correlated to the slip length and velocity, i.e. an increase in 
slip length and velocity corresponds to a larger drag 
(instead of reduction as in the cases with 2 fluids and the interface). This is because 
the changes to the turbulence structure and secondary motion, $O(\epsilon)$, overcome 
the contribution due to the slip length or velocity. In fact,
following \cite{Rastegari2015}, $O(\epsilon)$ is dominated by the Reynolds stress. 
In the case of staggered cubes with no interface, the Reynolds stress is much larger 
than in the case with the interface (Fig.\ref{fig:dudy-uv}a) thus explaining the difference with respect to the model based on the slip length.
\begin{figure}
\begin{center}
	\psfrag{xlab}{\relsize{+2} $\mu d\overline{U}/dy, \overline{uv}$}
\psfrag{ylab}{\relsize{+2} $y/h$}
	a)
\resizebox*{0.45\textwidth}{!}{\includegraphics{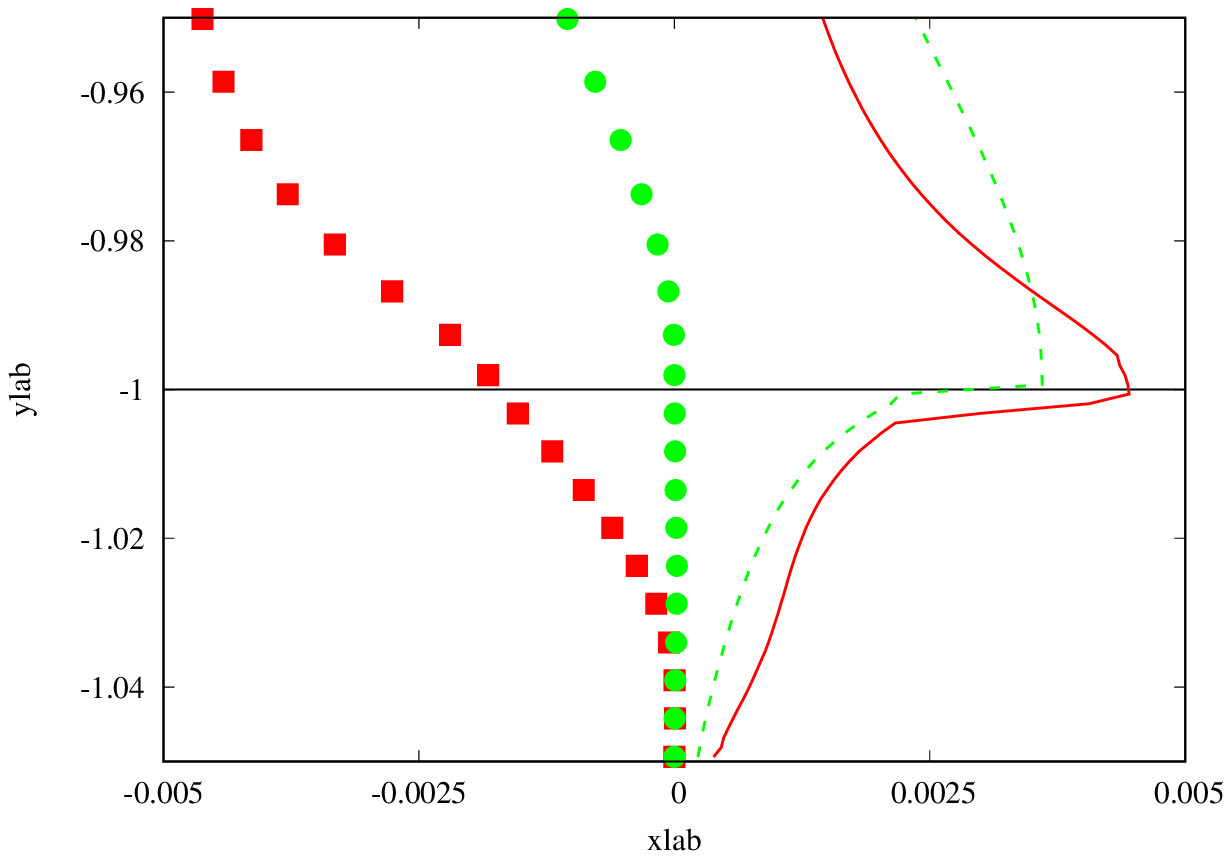}} 
	b)
\resizebox*{0.45\textwidth}{!}{\includegraphics{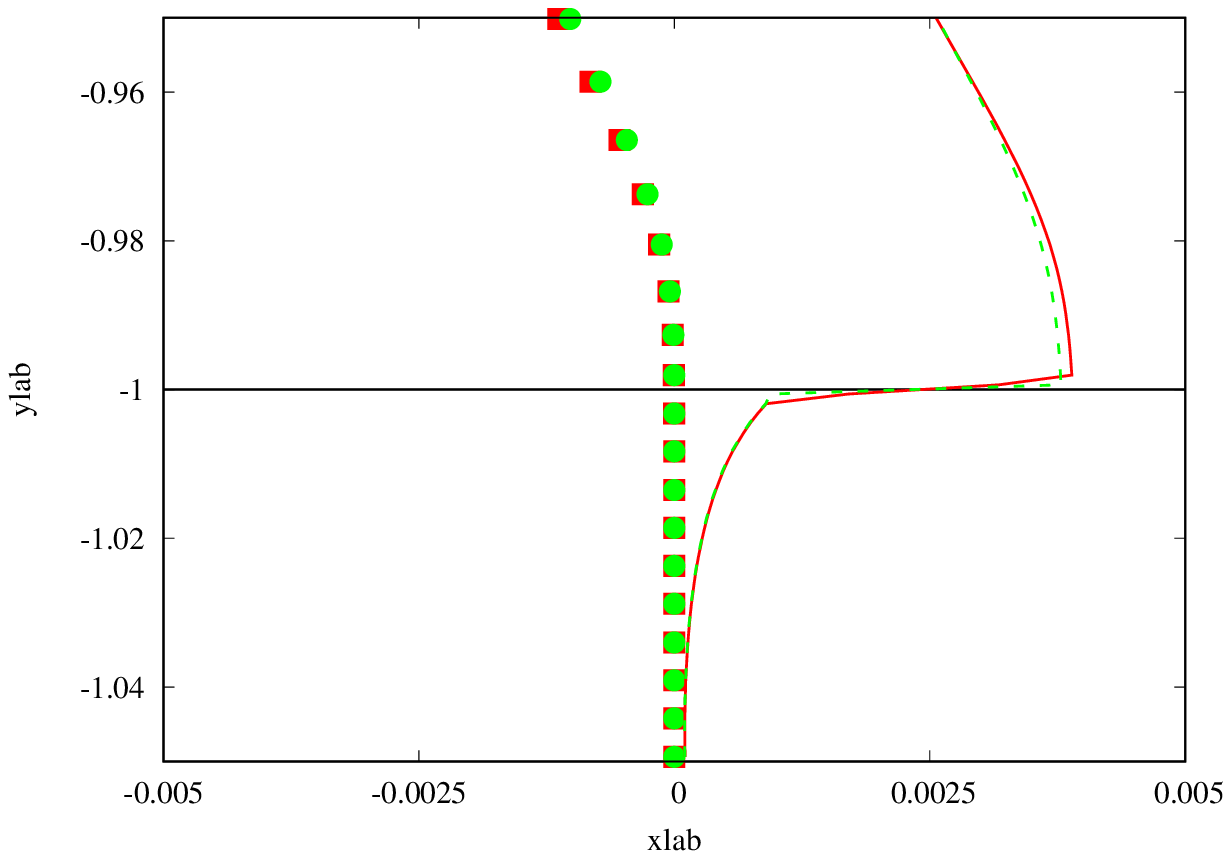}} 
\end{center}
	\caption{ Shear stress (lines) and Reynolds stress (symbols) near the interface ($y/h=-1$) 
	for staggered cubes (a) and longitudinal bars (b):  
	 ({\color{red} $\blacksquare$, \solid})one fluid only and 
	 ({\color{green} $\bullet$, \dashed}) LIS with $N=1$. Points are plotted every 4 for clarity.
}
\label{fig:dudy-uv}
\end{figure}
On the other hand,  longitudinal bars with a single phase agree better with the model because 
the Reynolds stress is very close to that obtained with two superposed fluids with an interface
 (Fig.\ref{fig:dudy-uv}b).
In case of a single fluid,  $\overline{uv}$ increases significantly 
near the leading edge of obstacles perpendicular to the flow direction. Longitudinal 
square riblets do not present walls orthogonal to the flow direction and therefore 
$\overline{uv}$ on the crests plane is
much smaller than that over staggered cubes.
In fact, as shown in table 1, this is the case where the interface has the weakest effect in terms
of drag, i.e. the difference of the drag relative to the cases with and without interface is only
$3\%$ while for staggered cubes it is about $70\%$.
}

{ 
To clarify why the interface contributes to reduce the drag, 
we integrated the time averaged Navier-Stokes equations for $U$ 
in the texture (the volume $V_t$ between the interface, bottom wall and side walls of the cavities):

\begin{equation}
\int_{V_t} \left( \frac{\partial \langle U \rangle}{\partial t}
+\frac{\partial{\langle U U_{j} \rangle}}{\partial{x_{j}}} \right) dv
=\int_{V_t} \left( -\frac{\partial \langle P \rangle }{\partial x}+
\frac{1}{Re_1} \frac{\partial^{2}\langle U \rangle }{\partial{x_{j}^{2}}} + \langle  \Pi \rangle
\right) dv \; .
\label{eq:texture1}
\end{equation}
Similarly to the previous derivation,
the flow rate is constant ($\int_{V_t} \partial{\langle U \rangle }/\partial{t}=0$) and because of the  
 periodicity in $x$ and $z$ and no-slip condition at the walls, Eq. \ref{eq:texture1}
 can be simplified to
\begin{equation}
	\int_{S_{int}}  \frac{1}{Re} \frac{d \langle U \rangle}{d y}|_{y_{int}} - \langle uv \rangle |_{y_{int}} ds
	+ \langle \Pi \rangle V_t
	= \int_{S_b} \langle P \rangle \vec{n} \cdot \vec{x} ds
	+\int_{S_{cav}}  \frac{1}{Re_1} \frac{d \langle U \rangle}{d n}|_{cav}  \vec{x} \cdot \vec{ds} \; ,
\label{eq:texture2}
\end{equation}
where $y_{int}$ is the location of the interface, $S_{int}$ is the area between the two fluids.
Equation \ref{eq:texture2} is a balance between the sum of   shear stress and Reynolds stress at the interface (total stress) 
and the form and frictional drag inside the cavity.

{ 
SHS and LIS reduce the velocity gradient at the crests plane of the texture. 
However,  to some approximation this also occurs in the case of one fluid only 
(i.e. rough wall). The shear stress relative to staggered cubes, with and without interface is only 
slightly different (Fig.\ref{fig:dudy-uv} solid lines).
On the other hand, the Reynolds stress is much larger in the case without interface
(Fig.\ref{fig:dudy-uv} symbols).
Consistent with previous studies on rough walls (\cite{Leonardi2003} and 
\cite{Leonardi2015}),
 the textures with one fluid only (and no interface) have a very large Reynolds stress at the crests plane
due mostly to
its dispersive component $\overline{\tilde{U}\tilde{V}}$ as discussed in 
\cite{Jelly2018}.
Therefore, 
the major difference between rough walls and either LIS or SHS is the magnitude
of  $\langle uv \rangle|_{y_{int}}$ at the interface.
The drag breakdown 
is approximately  the same, 
as it was observed in Fig. 5, implying that the fraction of the total drag due to
the cavities is the same in both cases.
Since $\langle uv \rangle|_{y_{int}}$ is smaller over SHS/LIS 
(zero in the ideal case of flat and slippery interface considered here) 
 the drag contribution due to the cavities and then the overall drag
is smaller too.

This explains why the same texture can decrease the drag with a fluid-fluid interface 
but increases the drag without an interface.
The cavities in both rough walls, SHS and LIS reduce the shear at the wall because of the slip velocity.
However, one must consider the extra contribution to the drag given by the texture. 
Over rough walls, it overcomes the reduction of the drag due to the slip
velocity and the total  drag increases. Because SHS and LIS present an interface which reduces the wall normal fluctuations and the momentum transport 
in the cavities, the contribution of the texture to the drag is much smaller 
 and, as a consequence, the total drag decreases. It is a combination of the slip velocity and the reduced turbulent transport due to the interface to 
 decrease the drag. 
}

{
The beneficial effect of the interface in reducing the drag can be also assessed 
by looking at the secondary streamwise vortices over longitudinal bars.
\begin{figure}
    \centering
    \psfrag{xlab}{$z/h$}
    \psfrag{ylab}[][][1][180]{$y/h$}
    \psfrag{c1}{}
a)    \includegraphics[width=0.4\linewidth]{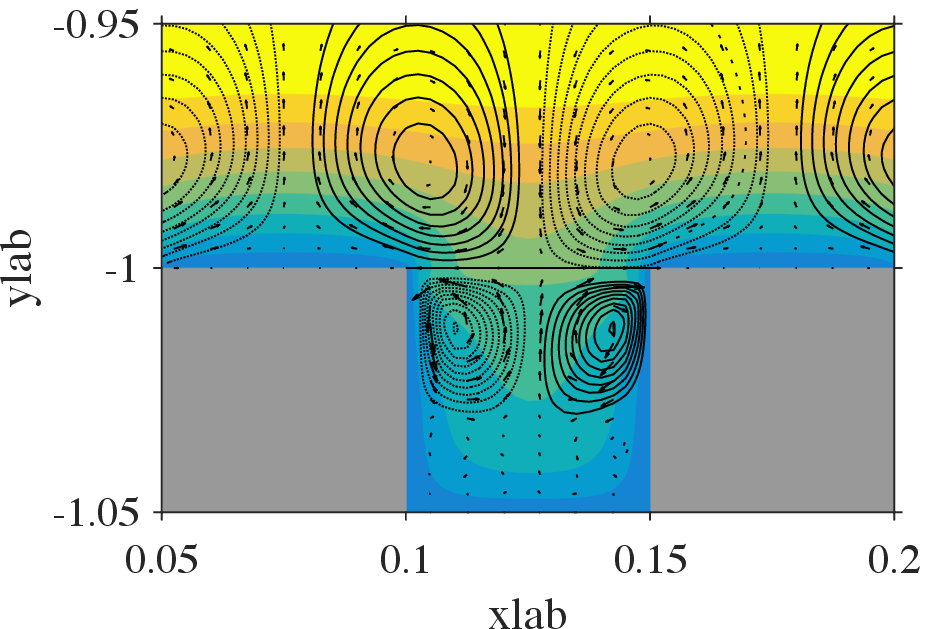} \\
b)    \includegraphics[width=0.4\linewidth]{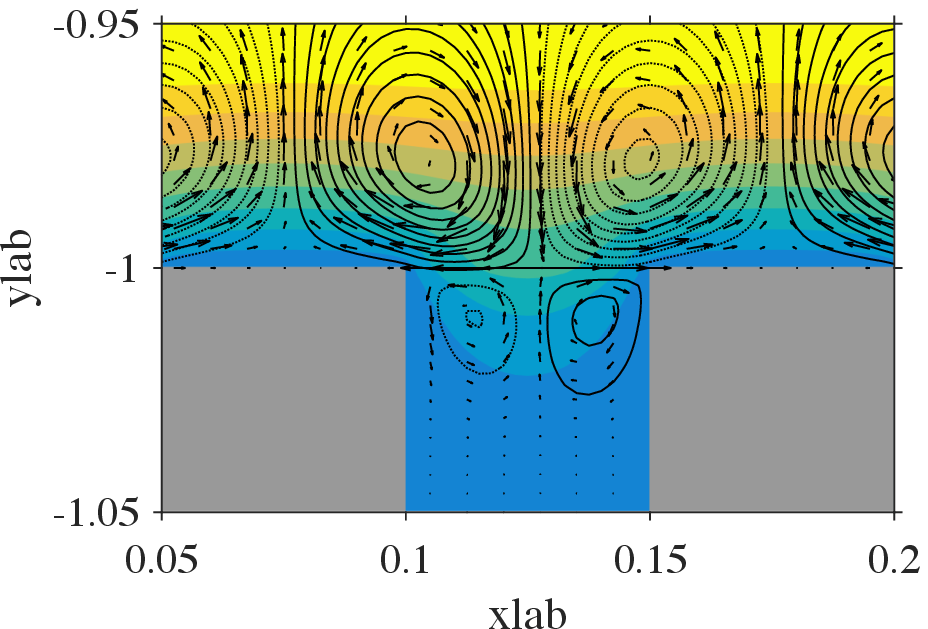}\\
c)     \includegraphics[width=0.4\linewidth]{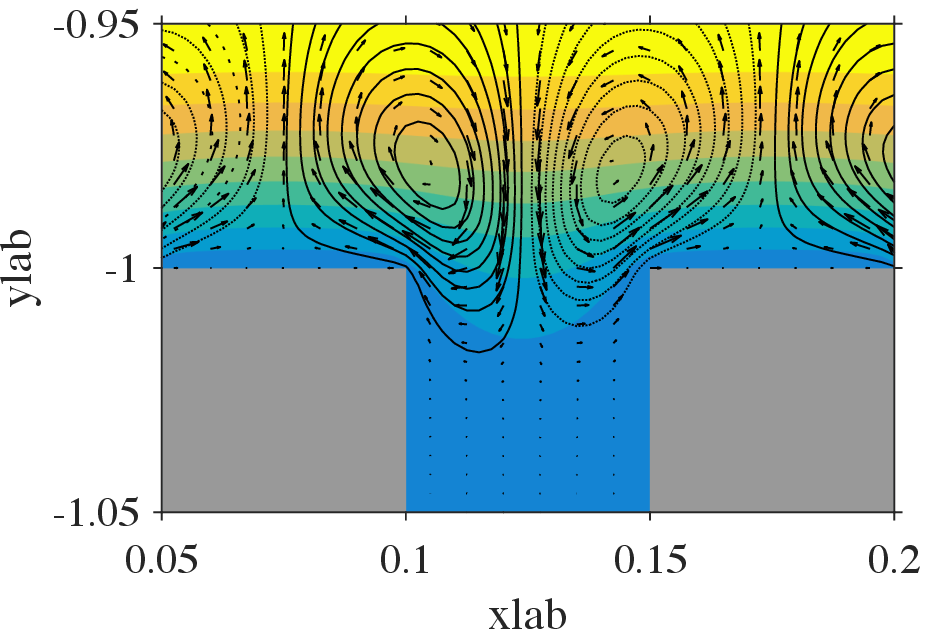}
    \caption{Streamlines of the time averaged secondary motion in the $z-y$ plane (solid clockwise and dotted counter-clockwise)
    superposed to color contours of time averaged streamwise velocity: a)  $N=100$, b) $N=2.5$, c) one fluid without interface. }
    \label{fig:secondary}
\end{figure}
A weak pair of streamwise vortices
near the riblet tip has been observed by  \cite{Goldstein}, \cite{Choi1993}, \cite{Suzuki1994}, 
and \cite{Crawford1996}.
 Such vortices cause a weak mean flow upward
the riblet crest and downward in the middle of the cavities.
They are the result of the vortex tilting $\omega_y d \overline{U}/dy$ which, in case of longitudinal bars, has a peak at the side walls.
In Fig. \ref{fig:secondary} the secondary motion for $N=100$, $N=2.5$ and for the case with a single 
fluid without interface are compared.
The interface keeps the streamwise vortices above the cavity while, for the case of single fluid, 
they penetrate below the crests plane. 
As a consequence, and 
consistently with the  discussion of Eq. \ref{eq:texture2}, the Reynolds stress 
$\langle uv \rangle|_{y_{int}}=0$ at the interface for $N=100$ and $2.5$ while 
 $\langle uv \rangle|_{y_{int}}\ne 0$ for the configuration without interface. 
 In fact, the cases with $N=100$ and $2.5$ reduce the drag while the case without
 interface does not.
The interface tends to damp the wall normal fluctuations and to keep the streamwise secondary vortices above the cavities. 
This reduces the momentum transfer inside the cavities and, therefore, the frictional and form drag of the texture. {
Classical studies of triangular riblets (one fluid without interface) showed that the cavity width had to be smaller than the size of the vortices to have drag 
reduction (\cite{Choi1993}, \cite{Mayoral2011})}. 
For SHS-LIS it is the interface keeping the vortices above the cavities thus reducing the Reynolds stress and then the drag. 
To some extent, SHS-LIS do not suffer from the geometrical limit of classical riblets, and  can achieve a very large theoretical drag reduction 
when the cavities are very large.
In reality the interface is deformable and not slippery as assumed in the present model.
\cite{Garcia2018} and \cite{Seo2018JFM} showed that when the interface deforms the amount of drag reduction is drastically reduced. 
Though exploring the role of interfacial deflections will be important to understanding many of these slippery surfaces, such an exploration is beyond the scope of this manuscript. However, the momentum transport associated with interfacial deflections will be captured by
Eq.  \ref{eq:texture2}. In fact, a deformation of the interface induces a Reynolds stress ($\langle uv \rangle|_{y,int}\ne 0$), an increased 
momentum transfer in the cavities and then a larger form and frictional drag of the texture.
{ Consequently, one might expect that decreasing the interfacial tension will lead to larger deformations of the interface, reducing the amount of drag reduction.
}
}

\section{Turbulent intensities }

\begin{figure}
\begin{center}
\psfrag{xlab1}{\relsize{+4} $y/h$}
	\psfrag{ylab2}[][][1][180]{\relsize{+4} $\sqrt{\overline{\tilde{U}\tilde{U}}},\sqrt{\overline{uu}}$ }
\subfigure[]{
\resizebox*{0.45\textwidth}{!}{\includegraphics{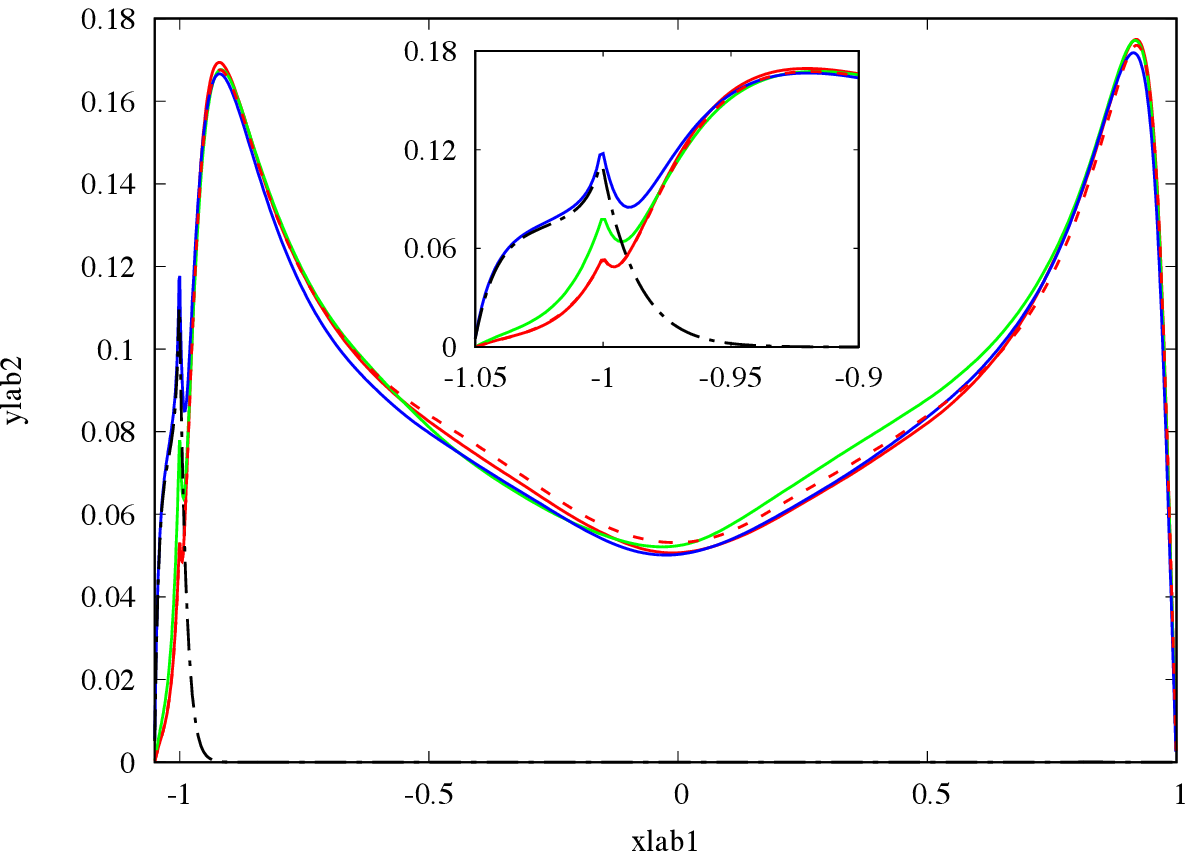}}
}
	\psfrag{ylab2}[][][1][180]{\relsize{+4} $\sqrt{\overline{\tilde{V}\tilde{V}}},\sqrt{\overline{vv}}$}
\subfigure[]{
\resizebox*{0.45\textwidth}{!}{\includegraphics{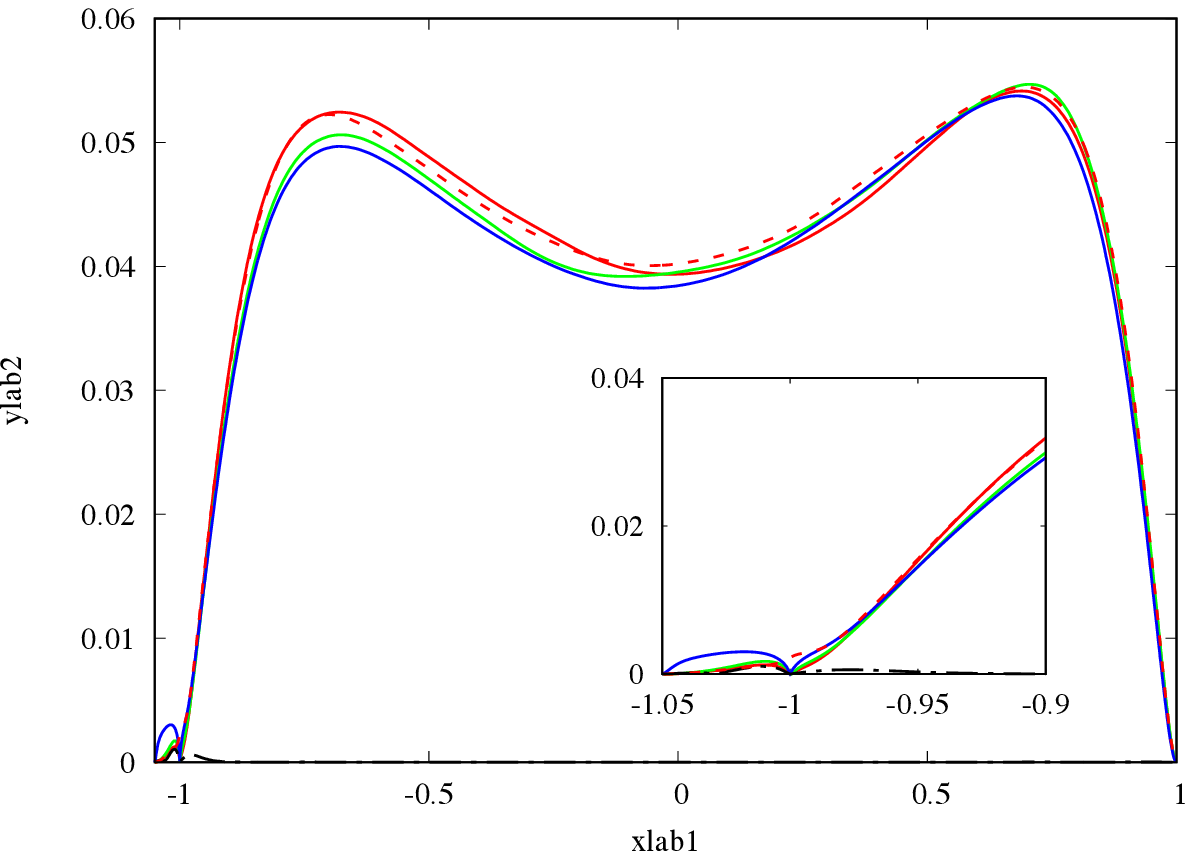}}
}
	\psfrag{ylab2}[][][1][180]{\relsize{+4} $\sqrt{\overline{\tilde{U}\tilde{U}}},\sqrt{\overline{uu}}$ }
\subfigure[]{
\resizebox*{0.45\textwidth}{!}{\includegraphics{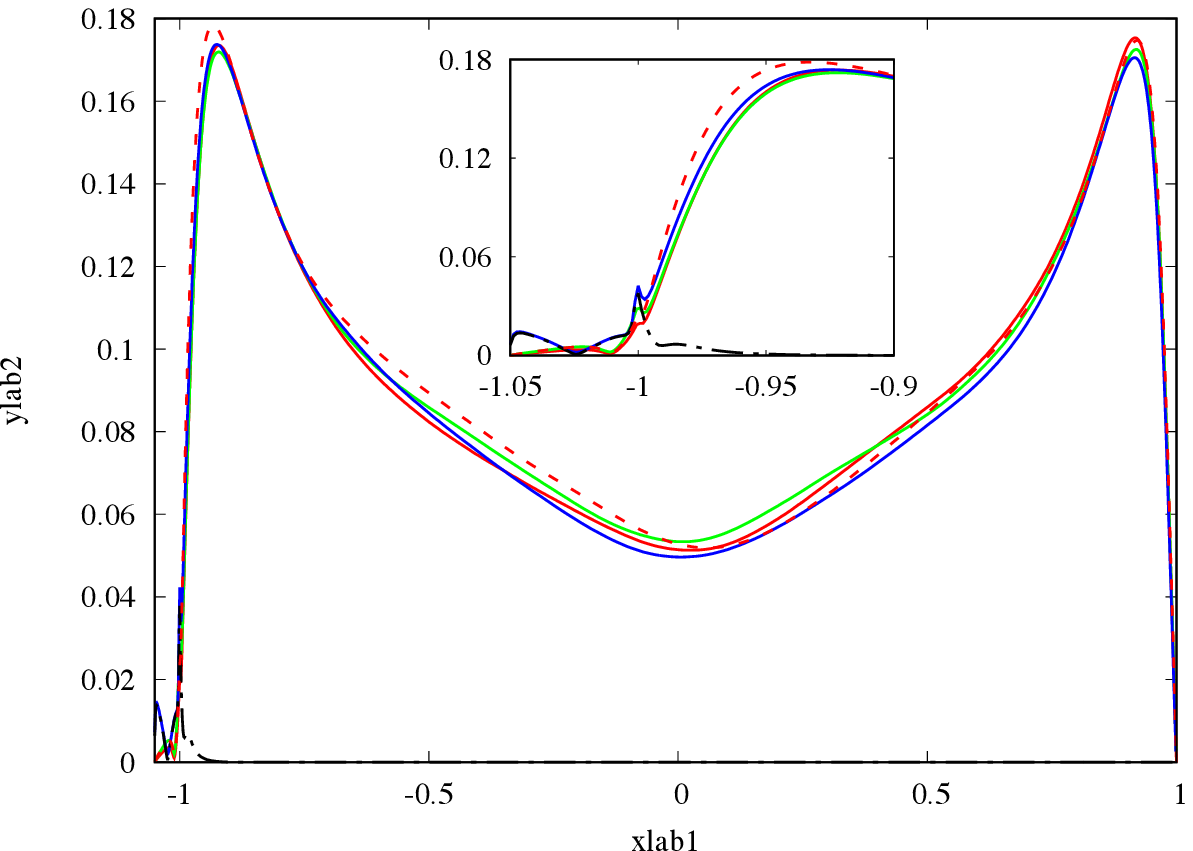}}
}
	\psfrag{ylab2}[][][1][180]{\relsize{+4} $\sqrt{\overline{\tilde{V}\tilde{V}}},\sqrt{\overline{vv}}$}
\subfigure[]{
\resizebox*{0.45\textwidth}{!}{\includegraphics{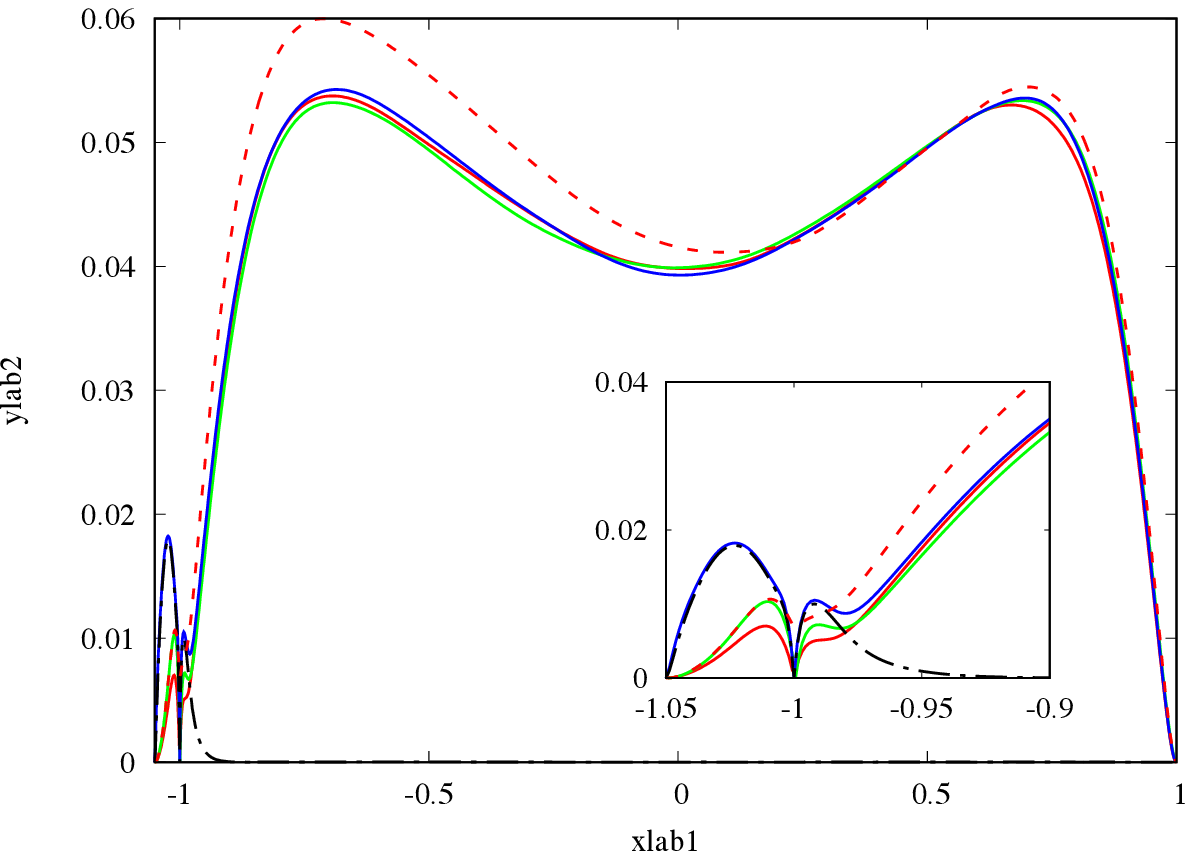}}
}
	\psfrag{ylab2}[][][1][180]{\relsize{+4} $\sqrt{\overline{\tilde{U}\tilde{U}}},\sqrt{\overline{uu}}$}
\subfigure[]{
\resizebox*{0.45\textwidth}{!}{\includegraphics{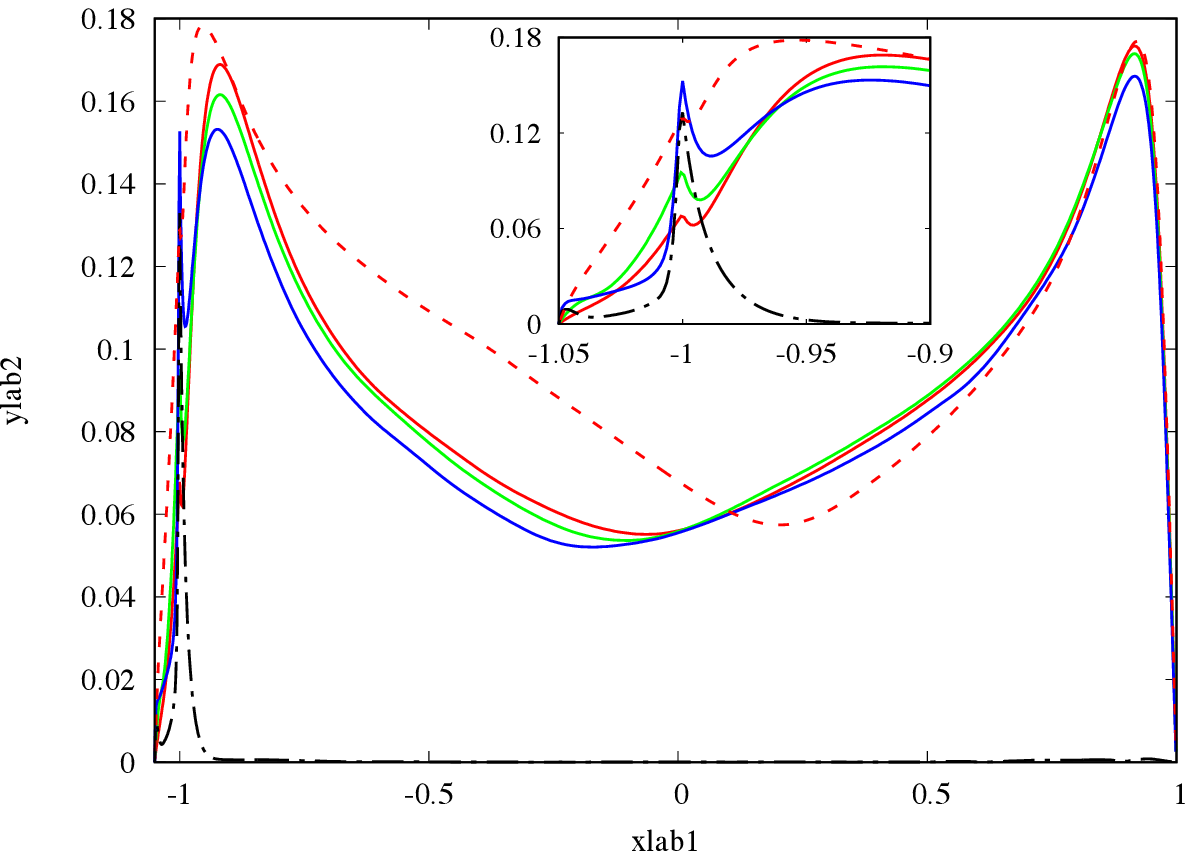}}
}
	\psfrag{ylab2}[][][1][180]{\relsize{+4} $\sqrt{\overline{\tilde{V}\tilde{V}}},\sqrt{\overline{vv}}$}
\subfigure[]{
\resizebox*{0.45\textwidth}{!}{\includegraphics{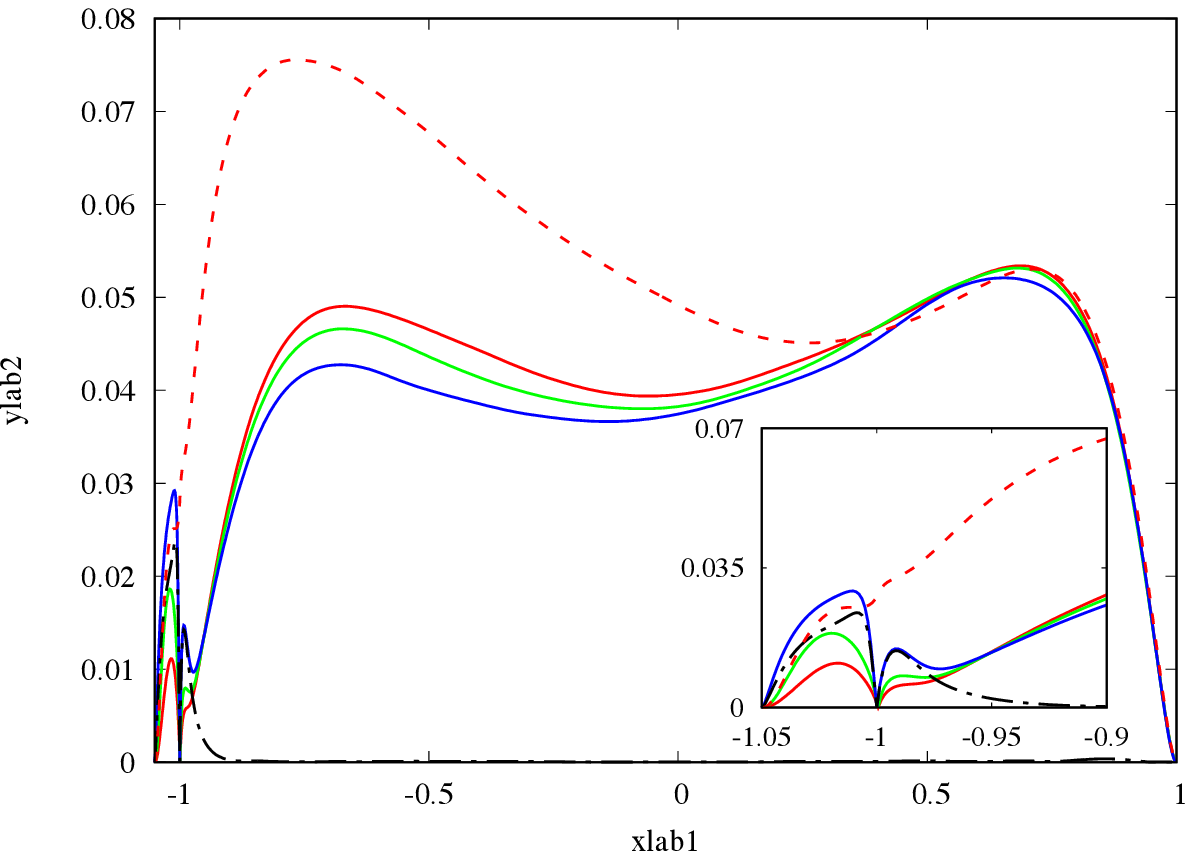}}
}
\caption{Root mean square of streamwise velocity (Left) $\sqrt{uu}$, and wall normal velocity (Right) $\sqrt{vv}$: \textbf{a),b)} Longitudinal
	Square Bars, \textbf{c),d)} Transversal Square Bars and \textbf{e),f)} Staggered Cubes $a=0.875$. 
	Dashed lines  one fluid configuration, solid lines for two superposed fluids,  {\color{red} \solid} $N=1$, {\color{green} \solid} $N=2.5$. For $N=100$ both the total stress  {\color{blue} \solid}, and its dispersive component {\color{black} \chndot} $\sqrt{\overline{\tilde{U_i}\tilde{U_j}}}$ are shown.
	A zoom near the texture is shown in the inset. { The velocity fluctuations are normalised with the bulk velocity}.
	}
\label{fig:uu_vv_N}
\end{center}
\end{figure}

The root mean square (rms) of streamwise and wall normal velocity fluctuations for longitudinal and transversal square bars and for 
staggered cubes with $a=0.875$ are shown in Figure \ref{fig:uu_vv_N} for $N=1,2.5,100$ and 
the case of one fluid only.
The upper smooth wall can be used as reference to assess the increase or reduction of the turbulent intensities.
The peak on the upper wall does not change much with $N$ indicating a weak correlation between the two walls.
On the lower wall, for the drag reducing configurations, such as longitudinal bars and staggered cubes with $a=0.875$, 
the wall normal and streamwise velocity fluctuations are reduced. The more the drag is reduced (the larger is $N$) the more the $rms$ decreases with respect to a smooth wall. In this cases,
the minimum of the velocity $rms$ is shifted towards the textured wall. As shown by \cite{Leonardi2005},
the position with respect to the centerline of the minimum of the  turbulent intensities is a measure of the relative contribution of the two walls to the total drag of the channel.

The peak of $\sqrt{\overline{uu}}$ inside the substrate (inset of Fig. \ref{fig:uu_vv_N}) is due to the dispersive component of the $rms$ due to inhomogeneities
of the time averaged velocity in spanwise and streamwise directions.
In fact, the time averaged
streamwise velocity  is very large above the cavities, zero on the crests 
of the cavities, and very low near the side walls thus inducing a
dispersive component that dominates near the interface between the two fluids
(this component is also present over regular rough walls). 
{ The dispersive component of the stress is significant in the region of the texture only, slightly above the crests plane it becomes very small (see {\color{black} \chndot} in the inset of
Fig. \ref{fig:uu_vv_N}). Similar trend of the dispersive stress was observed over rough walls in \cite{Leonardi2015}, and \cite{Leonardi2010}.
It was also shown that
dispersive stresses increase significantly when averages in time are done with a small number of data fields and the secondary vortices are locked in a position within the texture.
}

An increase of the velocity fluctuations is observed for the transversal square bars and the staggered cubes with one fluid.
This is consistent with results of the flow over rough surfaces. In this case, the minimum of the standard deviation  is shifted towards
the upper wall which is the wall with lowest drag.

\begin{figure}
\begin{center}
        \psfrag{xlab1}{$Re_\tau$}
	\psfrag{ylab2}[][][1][180]{$\overline{vv}_{max}^+$}
\includegraphics[width=0.45\textwidth]{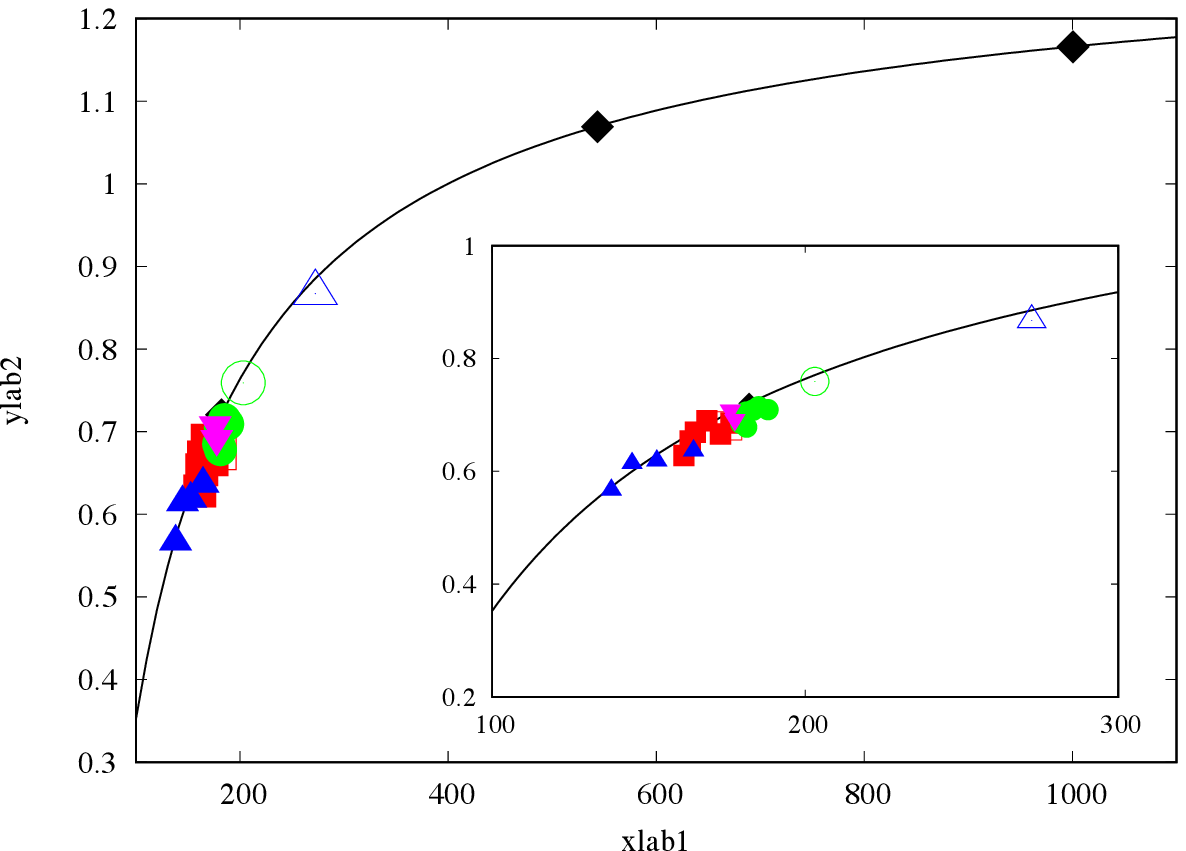}
	\psfrag{ylab2}[][][1][180]{$\overline{uu}_{max}^+$}
\includegraphics[width=0.45\textwidth]{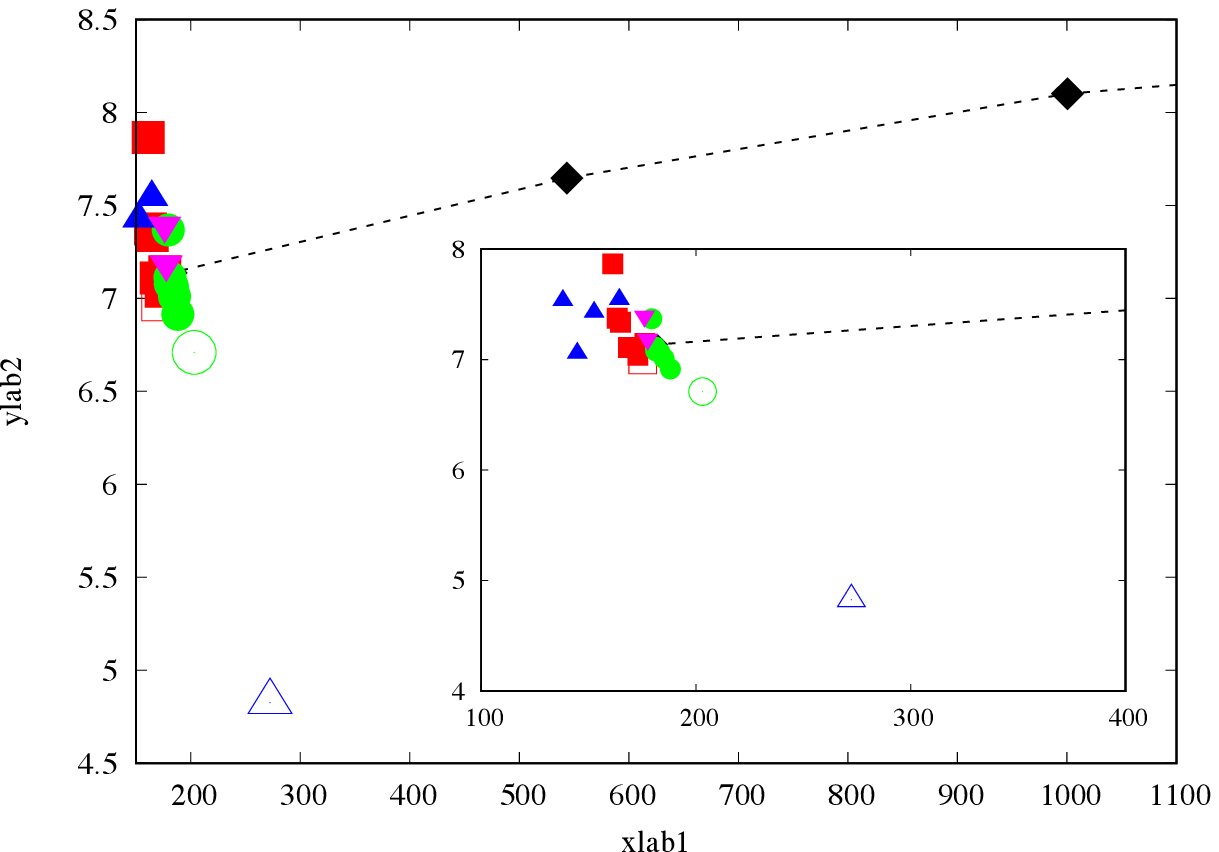}
\end{center}
	\caption{ Maximum wall normal (a) and streamwise (b) velocity fluctuation as function of $Re_\tau$,
{\color{red} $\blacksquare$} longitudinal bars; 
	{\color{green} $\bullet$} transversal square bars;
{\color{purple} $\blacktriangledown$} staggered cubes with $a=0.5$;
	 {\color{blue} $\blacktriangle$} staggered cubes with  $a=0.875$.
Empty symbols indicate simulations with only one fluid without interface,
	 solid symbols simulations with two fluids and a slippery interface,
	solid line  Eq. \ref{eq:vvmax}, dashed line and black diamonds smooth wall data from \citep{Lee2015}.}
\label{fig:uu-vvplus}
\end{figure}
{ 
The correlation between velocity fluctuations and wall shear stress is further corroborated 
by plotting the maximum of the wall normal velocity fluctuation, scaled in wall units, as function of the turbulent
Reynolds number $Re_\tau$ 
(Fig. \ref{fig:uu-vvplus}a).
{ 
For each case here considered, the turbulent Reynolds number has
been calculated as
$Re_\tau= U_\tau y_{uv}/\nu$  
where $U_\tau$ is the friction velocity on the lower wall (the one with the texture)
and  $y_{uv}$ is the distance from the texture to the  
zero crossing of  $\overline{uv}$ as suggested in \cite{Burattini}.
When the drag on the two walls is approximately the same, $y_{uv}$ is very close to the half height of the
channel. However, in the cases with higher drag reduction or drag increase, the shift of  $y_{uv}$ 
with respect to the centerline is significant and affects the value of $Re_\tau$.
}
Numerical results are compared with a best fitting of smooth wall data from \citep{Lee2015},
\begin{equation}
(vv)_{max}^+=A Re_\tau^C+B \;,	
	\label{eq:vvmax}
\end{equation}
with $A=-38.27$, $B=1.32$ and $C=-0.7986$.
Results from SHS-LIS and rough surfaces follow, to a close approximation, the trend of the 
smooth wall. A change in the peak of $\overline{vv}^+$ with respect to the baseline smooth wall at $Re_\tau=180$ 
is, therefore, primarily due to the overall change of shear and
can consequently be expressed as a function of $Re_\tau$ for that particular surface.
While the wall normal velocity fluctuations depend primarily on the overall shear, 
 the streamwise velocity fluctuations vary significantly
with respect to the smooth wall and a clear trend could not be determined (Fig.\ref{fig:uu-vvplus}b).
Previous studies on rough surfaces highlighted an increase of isotropy with respect to
the smooth wall. 
Here we add that $\overline{vv}$ increases proportionally to the shear and the turbulent Reynolds number,  while $\overline{uu}$ 
does not increase with the same rate. When scaled in wall units, $\overline{vv}^+_{max}$ is approximately the same as that of a smooth wall at the same turbulent
Reynolds number, while $\overline{uu}^+_{max}$ becomes much lower as in the case of roughness made of staggered cubes ({\color{blue} \tridot}  in Fig.\ref{fig:uu-vvplus}). 
}
\begin{figure}
\begin{center}
        \psfrag{xlab1}{$\max(\overline{vv})/\max(\overline{vv}_{0})$}
	\psfrag{xlab1}{$\overline{vv}_{max}/\overline{vv}_{max,0}$}
\psfrag{ylab2}{$1-\tau/\tau_0$}
\includegraphics[width=0.45\textwidth]{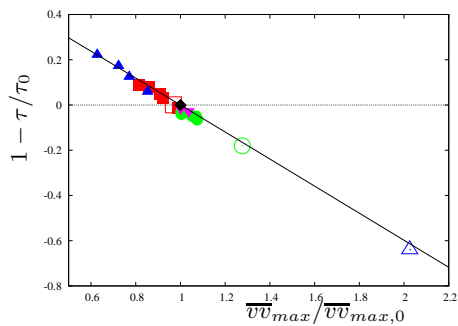}
\end{center}
	\caption{
Dependence of the amount of drag reduction  with the maximum of the wall normal velocity rms:
{\color{red} $\blacksquare$} longitudinal bars; 
	{\color{green} $\bullet$} transversal square bars;
{\color{magenta}  $\blacktriangledown$} staggered cubes with $a=0.5$;
	 {\color{blue} $\blacktriangle$} staggered cubes with  $a=0.875$;
	 solid line, Eq. \ref{eq:vvmax-fin}.
Empty symbols indicate simulations with only one fluid without interface,
	 solid symbols simulations with two fluids and a slippery interface.}
\label{fig:N_vv_dr}
\end{figure}

Similarly, we can derive a correlation between
the amount of drag reduction ($DR=1-\tau/\tau_0$)
and the
maximum of the wall normal velocity fluctuations (normalized with that of the smooth channel), 
$\overline{vv}_{max}/\overline{vv}_{max,0}$.
As shown in Fig.\ref{fig:N_vv_dr},
regardless of the
layout of the substrate (cubes, longitudinal or transversal square bars),
 viscosity ratio, and whether we have one or two fluids separated by an interface, 
 there is a strong correlation between the wall shear stress and 
$\overline{vv}_{max}/\overline{vv}_{max,0}$.
A larger maximum of wall normal velocity fluctuations
corresponds to larger drag. To our best knowledge,  this is the first time
super--hydrophobic, liquid--infused surface, and rough surfaces are reconciled under the same
scaling {(while correlations between the drag and the slip length or velocity work well only for SHS and LIS with weak secondary motion and Reynolds stresses)}.
This confirms that wall shear stress and wall normal velocity fluctuations are 
strongly tied, despite the different boundary condition at the wall (smooth, 
made of cavities filled with the same or another fluid, with or without an interface).

{ 
The correlation can be derived from  Eq.\ref{eq:vvmax}, which can be expressed as:
\begin{equation}
	\frac{vv_{max}}{U^2_\tau}=A Re_\tau^C+B \;,
	\label{eq:1}
\end{equation}
Similarly for the baseline smooth wall case at $Re_{\tau,0}=180$:
\begin{equation}
	\frac{vv_{0,max}}{U^2_{\tau,0}}=A Re_{\tau,0}^C+B \;,
	\label{eq:2}
\end{equation}
Dividing Eq.\ref{eq:1} by Eq. \ref{eq:2}, we obtain
\begin{equation}
	\frac{vv_{max}}{vv_{0,max}}=	\frac{U^2_\tau}{U^2_{\tau,0}} \frac{A Re_\tau^C+B}{A Re_{\tau,0}^C+B}
	\label{eq:3}
\end{equation}
 $Re_\tau$ can be expressed as 
\begin{equation}
	Re_\tau=Re_{\tau,0} \frac{y_{uv}}{h} \frac{U_\tau}{U_{\tau,0}}.
\end{equation}
 The value of $y_{uv}/h$ can be derived from \cite{Leonardi2005}
 neglecting the volume of the texture with respect to the volume of the channel:
\begin{equation}
	\frac{y_{uv}}{h} \simeq  \frac{2}{1+(U_{\tau,0}/U_\tau)^2},
\end{equation}
 therefore
\begin{equation}
	\frac{vv_{max}}{vv_{0,max}}=	\frac{U^2_\tau}{U^2_{\tau,0}} 
	         \frac{A (\frac{U_\tau}{U_{\tau,0}}
		 \frac{2}{1+(U_{\tau,0}/U_\tau)^2})^C  Re_{\tau,0}^C+B}{A Re_{\tau,0}^C+B}  \;.
\end{equation}
Substituting $U_\tau=\sqrt{\tau/\rho}$,
\begin{equation}
	\frac{vv_{max}}{vv_{0,max}}=	\frac{\tau}{{\tau_0}} 
	              \frac{A (\frac{\tau}{\tau_0})^{C/2}  
		      (\frac{2}{1+(\tau_0/\tau)})^C
		      Re_{\tau,0}^C+B}{A Re_{\tau,0}^C+B}  \;.
	\label{eq:vvmax-fin}
\end{equation}
Eq. \ref{eq:vvmax-fin} has been 
developed assuming only that the wall normal fluctuations over any texture
change as an effect of the change of
$Re_\tau$  using a correlation developed for smooth walls (Eq.6.1).
Eq. \ref{eq:vvmax-fin} agrees well
with numerical results for SHS-LIS and rough walls as shown in Fig. \ref{fig:N_vv_dr}. This
implies that the overall shear is 
the main mechanism producing and sustaining  turbulence, regardless if it is produced 
in a particular texture, with or without an interface. 
Sufficiently above the texture, where the variability  of the flow due to spatial inhomogeneities is negligible,
the flow over different surfaces is the same as that over  a flat wall 
at the same $Re_\tau$. 
} {Equation \ref{eq:vvmax-fin} and Fig.\ref{fig:N_vv_dr} emphasize further that the role of the interface in damping the turbulent 
transport and the wall normal velocity fluctuations near the wall is critical to achieve drag reduction.}

\section{Conclusions}

Direct Numerical Simulations of two superposed (immiscible) fluids in a turbulent channel have been performed where one of the fluids is fully wetted within periodic substrate textures.
Various viscosity ratios  were evaluated for two-dimensional (longitudinal and transverse bars) and three-dimensional (cubic pillar) substrate patterns.

Staggered cubes with a large fluid-area fraction ($a=0.875$) and longitudinal square bars present the highest drag reduction which increases by decreasing the  viscosity of the fluid inside the substrate. 
Transverse square bars and staggered cubes with $a=0.5$, for this particular pitch to height ratio, increase drag overall
as the form drag negates the positive effects of a reduced shear stress above the cavities. 

DNS results showed that slip length, slip velocity, frictional and form drag and amount of drag reduction
vary smoothly with $N$   indicating that SHS and LIS reduce the drag
with a similar mechanism: 
the flow inside the substrate reduces the shear of the main stream above the cavities through a slip velocity.

Perhaps surprising in the case of the LIS, drag reduction can be achieved  with fluid of a viscosity similar to that of 
the bulk flow $N \simeq 1$. To better understand this result, an additional set of simulations with the same surface structures with a single fluid only has been carried out.
By comparing the results for a single fluid and two fluids separated by an interface, 
with the same substrate,
it was shown that the interface plays a key role in reducing the drag.
The interface damps wall normal velocity fluctuations and then limits the flux of momentum inside the cavities.
{ 
The slip velocity, slip length and the viscous shear above 
a rough wall and a LIS with $N=1$ made with the same texture are very similar.
However, the amount of momentum dissipated
in the cavities is significantly lower for the LIS because of 
 the reduced Reynolds stress (zero in case of ideal slippery interface) at the interface. This 
explains why a LIS, despite being very similar to a 
classical rough wall, reduces the drag.}
It is not a particular shape of the textured surface which makes LIS  different from rough surfaces, but it is 
the interface between the two fluids preventing a momentum flux in the cavities.

The standard deviation of the velocity fluctuations is closely connected to the wall shear stress with a strong correlation between the amount of drag reduction and the maximum of 
the wall normal velocity fluctuations. This was shown for rough, super-hydrophobic and liquid infused surfaces, regardless of the viscosity 
ratio, shape of the substrate and  presence of the interface (recall that for rough walls the drag increases).

{ Although it is beyond the scope of this paper, while in this study we considered 
the interface slippery but not deformable in the vertical direction, in reality
the interface can deform. In this case, depending on the Weber number, and how 
the deformation of the interface is correlated with streamwise velocity fluctuations,
the momentum transfer inside the texture may be only partially reduced with 
a consequent detrimental effect on the drag. 
The amount of drag reduction is less than that obtained with a flat and slippery interface 
as shown by \cite{Garcia2018}.
In addition, more studies are needed to
optimize  the morphology of the texture and size of the cavities 
to avoid the depletion of the fluid and increase durability.}



\section*{Acknowledgments}
This research was supported  by  ONR MURI grants N00014-12-01-0875 and N00014-12-01-0962, program manager Dr. Ki-Han Kim. Matthew K. Fu was supported by the Department of Defense (DoD) through the National Defense Science \& Engineering Graduate Fellowship (NDSEG) Program. 
Numerical simulations were performed on TACC. Prof. Akhavan is acknowledged
for useful discussions.

\section*{Appendix}
{ 

\begin{figure}
    \centering
	    \psfrag{xlab}{$y/h$}
    \psfrag{ylab}[][][1][180]{$\overline{U}$}
    \psfrag{c1}{(a)}
    \includegraphics[width=0.45\linewidth]{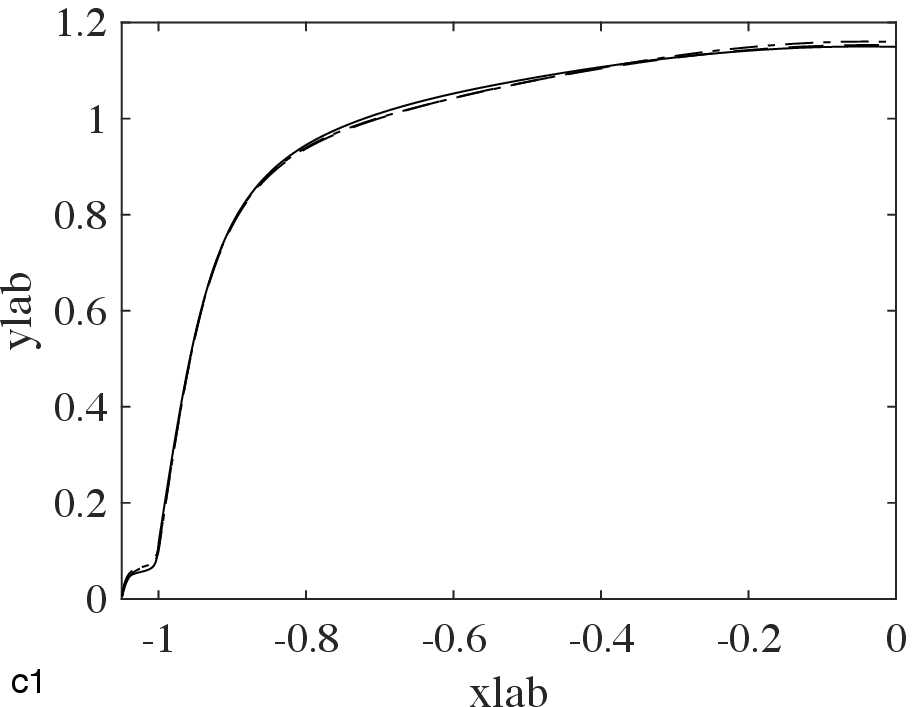}
    \psfrag{xlab}{$y/h$}
    \psfrag{ylab}[][][1][180]{}
    \psfrag{c1}{(b)}
    \psfrag{c2}{$\sqrt{\overline{u}\overline{u}}$}
    \psfrag{c3}{$\sqrt{\overline{w}\overline{w}}$}
    \psfrag{c4}{$\sqrt{\overline{v}\overline{v}}$}
    \includegraphics[width=0.45\linewidth]{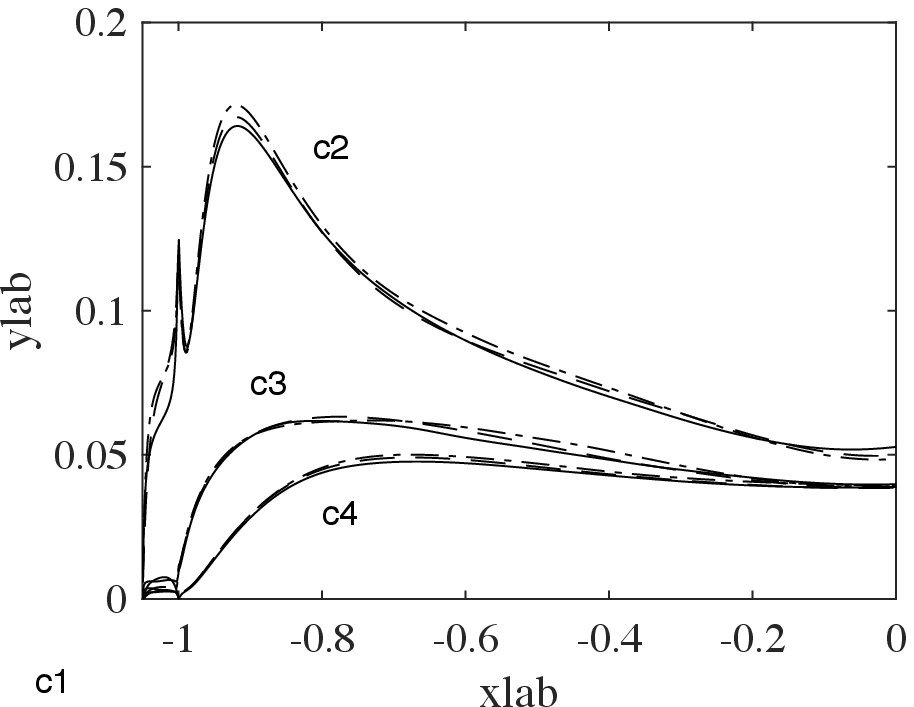}
    \psfrag{xlab}{$y/h$}
    \psfrag{ylab}[][][1][180]{$\overline{U}$}
    \psfrag{c1}{(c)}
    \includegraphics[width=0.45\linewidth]{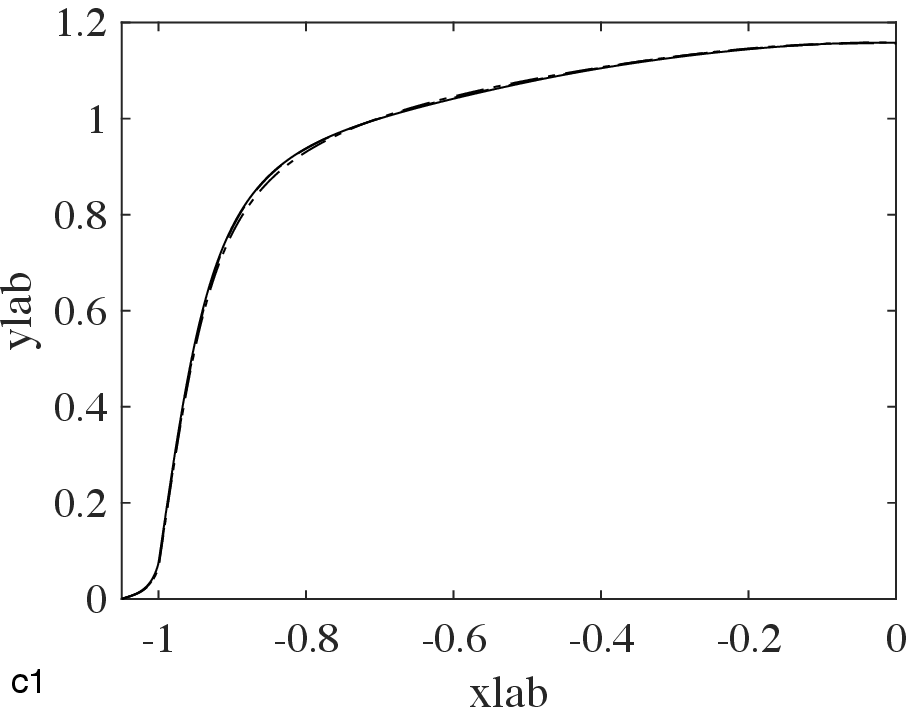}
    \psfrag{xlab}{$y/h$}
    \psfrag{ylab}[][][1][180]{}
    \psfrag{c1}{(d)}
    \psfrag{c2}{$\sqrt{\overline{u}\overline{u}}$}
    \psfrag{c3}{$\sqrt{\overline{w}\overline{w}}$}
    \psfrag{c4}{$\sqrt{\overline{v}\overline{v}}$}
    \includegraphics[width=0.45\linewidth]{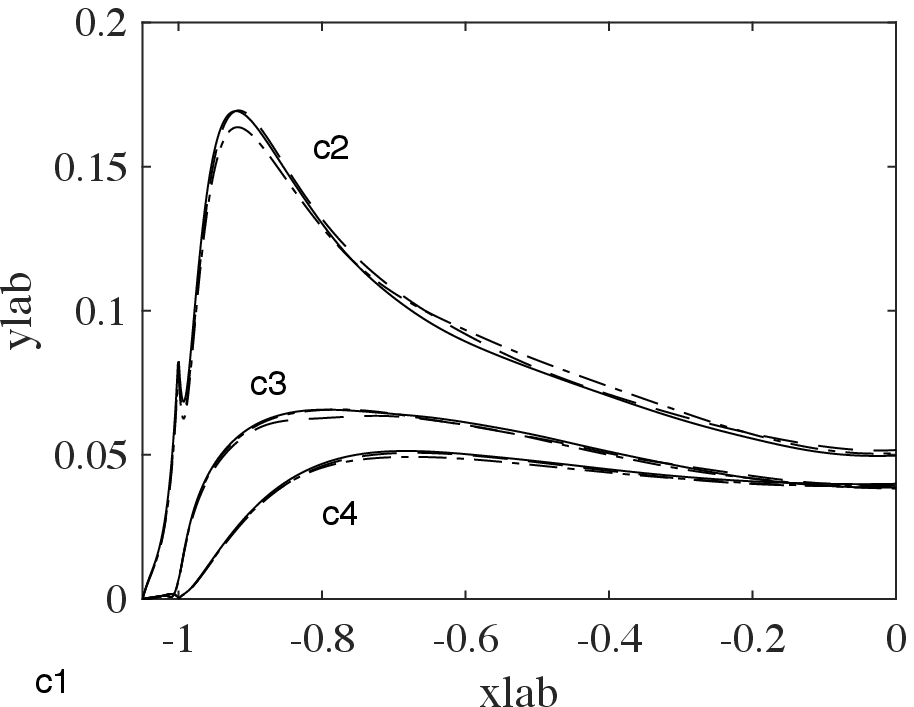}
	\caption{(a,c) Streamwise velocity profile and (b,d) rms of velocity fluctuations for longitudinal bars with viscosity ratio $N=100$ (a,b) and $2.5$ (c,d):
	 512x384x640 ($\solid$); 512x384x1280 ($\dashed$); 256x384x640 ($\dashdot$).  Velocities are normalised by the bulk velocity.}
    \label{fig:stat_transv}
\end{figure}
\begin{table}
    \centering
    \caption{Turbulent Reynolds number $Re_\tau$ and grid resolution 
	for the different cases considered in the paper: LSB, longitudinal bars;
	TSB, transversal bars; SC, staggered cubes; NST indicates a single fluid, $a$ the gas fraction and $N$ the viscosity ratio.
	A grid sensitivity analysis of the drag on the textured wall is also reported for longitudinal and transversal bars with $N=2.5$ and $100$.
	    }
\begin{tabular}{ccccccccccccc}
\hline
	Case	&	$N$	&	$a$	&	$Re_\tau$	&	$N_x$	&	$N_y$	&	$N_z$	&	$\Delta x^+$	&	$\Delta y^+_{min}$	&	$\Delta y^+_{max}$	&	$\Delta z^+$	&	Grid id	&	$\frac{(\tau_1-\tau_i)}{\tau_1}$ 	\\ \hline
LSB	&	0.1	&	0.5	&	178.4	&	512	&	384	&	1280	&	2.23	&	0.23	&	1.90	&	0.45	&	1	&	-	\\
LSB	&	1	&	0.5	&	174.6	&	512	&	384	&	1280	&	2.18	&	0.22	&	1.86	&	0.44	&	1	&	-	\\
LSB NST	&	1	&	0.5	&	177.2	&	512	&	384	&	1280	&	2.21	&	0.23	&	1.89	&	0.44	&	1	&	-	\\
LSB	&	2.5	&	0.5	&	173.1	&	512	&	384	&	1280	&	2.16	&	0.22	&	1.85	&	0.43	&	1	&	0.00\%	\\
LSB	&	2.5	&	0.5	&	174.2	&	512	&	384	&	640	&	2.18	&	0.22	&	1.86	&	0.87	&	2	&	0.48\%	\\
LSB	&	2.5	&	0.5	&	173.0	&	256	&	384	&	640	&	4.33	&	0.22	&	1.85	&	0.87	&	3	&	1.77\%	\\
LSB	&	10	&	0.5	&	172.4	&	512	&	384	&	1280	&	2.15	&	0.22	&	1.84	&	0.43	&	1	&	-	\\
LSB	&	20	&	0.5	&	170.1	&	512	&	384	&	1280	&	2.13	&	0.22	&	1.81	&	0.43	&	1	&	-	\\
LSB	&	100	&	0.5	&	169.3	&	512	&	384	&	1280	&	2.12	&	0.22	&	1.81	&	0.42	&	1	&	0.00\%	\\
LSB	&	100	&	0.5	&	169.0	&	512	&	384	&	640	&	2.11	&	0.22	&	1.80	&	0.85	&	2	&	0.37\%	\\
LSB	&	100	&	0.5	&	172.6	&	256	&	384	&	640	&	4.32	&	0.22	&	1.84	&	0.86	&	3	&	-3.90\%	\\
TSB	&	0.1	&	0.5	&	179.1	&	1280	&	384	&	512	&	0.90	&	0.23	&	1.91	&	1.12	&	1	&	-	\\
TSB	&	1	&	0.5	&	179.9	&	1280	&	384	&	512	&	0.90	&	0.23	&	1.92	&	1.12	&	1	&	-	\\
TSB NST	&	1	&	0.5	&	192.6	&	1280	&	384	&	512	&	0.96	&	0.25	&	2.05	&	1.20	&	1	&	-	\\
TSB	&	2.5	&	0.5	&	180.7	&	1280	&	384	&	512	&	0.90	&	0.23	&	1.93	&	1.13	&	1	&	0.00\%	\\
TSB	&2.5	&	0.5	&	182.1	&	640	&	384	&	512	&	1.82	&	0.23	&	1.94	&	1.14	&	2	&	-0.43\%	\\
TSB	&2.5	&	0.5	&	183.1	&	640	&	384	&	256	&	1.83	&	0.23	&	1.95	&	2.29	&	3	&	-1.45\%	\\
TSB	&	10	&	0.5	&	181.9	&	1280	&	384	&	512	&	0.91	&	0.23	&	1.94	&	1.14	&	1	&	-	\\
TSB	&	20	&	0.5	&	181.8	&	1280	&	384	&	512	&	0.91	&	0.23	&	1.94	&	1.14	&	1	&	-	\\
TSB	&	100	&	0.5	&	182.8	&	1280	&	384	&	512	&	0.91	&	0.23	&	1.95	&	1.14	&	1	&	0.00\%	\\
TSB	&100	&	0.5	&	182.0	&	640	&	384	&	512	&	1.82	&	0.23	&	1.94	&	1.14	&	2	&	0.93\%	\\
TSB	&100	&	0.5	&	184.7	&	640	&	384	&	256	&	1.85	&	0.24	&	1.97	&	2.31	&	3	&	-2.06\%	\\
SC	&	1	&	0.875	&	172.0	&	1280	&	384	&	640	&	0.86	&	0.22	&	1.83	&	0.86	&	1	&	-	\\
SC NST	&	1	&	0.875	&	227.0	&	1280	&	384	&	640	&	1.13	&	0.29	&	2.42	&	1.13	&	1	&	-	\\
SC	&	2.5	&	0.875	&	165.9	&	1280	&	384	&	640	&	0.83	&	0.21	&	1.77	&	0.83	&	1	&	-	\\
SC	&	10	&	0.875	&	161.2	&	1280	&	384	&	640	&	0.81	&	0.21	&	1.72	&	0.81	&	1	&	-	\\
SC	&	100	&	0.875	&	156.4	&	1280	&	384	&	640	&	0.78	&	0.20	&	1.67	&	0.78	&	1	&	-	\\
SC	&	2.5	&	0.5	&	178.6	&	1280	&	384	&	640	&	0.89	&	0.23	&	1.91	&	0.89	&	1	&	-	\\
SC	&	100	&	0.5	&	179.6	&	1280	&	384	&	640	&	0.90	&	0.23	&	1.92	&	0.90	&	1	&	-	\\
Channel	&	-	&	-	&	177.2	&	512	&	384	&	256	&	2.22	&	0.23	&	1.89	&	2.22	&	1	&	-	\\
\end{tabular}
    \label{tab:grref}
\end{table}
The computational box is $6.4h \times 2.05 h \times 3.2h$ in streamwise, wall-normal and spanwise direction as mentioned in Section 2.
The grid used for  transversal square bars is $1280 \times 384 \times  512$ in streamwise, wall normal and spanwise
direction, with a resolution of about $1$ wall units in both homogeneous directions. 
For staggered cubes, the number of points in spanwise direction was increased to $640$ in order to have at least $10$ points within the spanwise side of the cube.
On the other hand, for longitudinal bars
the grid is $512 \times 384 \times  1280$ in streamwise, wall normal and spanwise direction respectively. 
The increased resolution in spanwise direction is aimed at resolving the high velocity gradient at the edge of the 
longitudinal bars, as suggested by \citet{Jelly2014}.
A non--uniform grid is used in the wall normal direction with $40$ points clustered within the textured substrate. The vertical grid is the same for all textures.
The present grid is finer than that used in our previous work on roughness (\cite{Leonardi2003}, \cite{Leonardi2010}, \cite{Leonardi2003}, \cite{Burattini}).
A grid sensitivity  analysis has been performed to assess the dependence of the results on the resolution.
Details are in Table \ref{tab:grref}.
For longitudinal bars by halving the number of points in streamwise direction, the error in the friction coefficient is about $0.37 \%$ and $0.48 \%$ for $N=100$ and $2.5$ respectively. By halving the number of points also in spanwise direction the error increases to $3.9\%$ and $1.77 \%$ for 
$N=100$ and $2.5$ respectively. The resolution is less critical for smaller values of $N$ because the 
velocity gradient at the interface and at the side walls is smaller.
A similar trend is observed for transversal square bars.

Mean velocity profiles  and turbulent intensities are shown in Fig. \ref{fig:stat_transv} for $N=2.5$ and $100$. Results depend very weakly on the grid used and are hardly discernible. A larger sensitivity to the grid is observed  in  the region inside the cavities ($-1.05<y/h<-1$) for the case $N=100$ where the velocity gradients at the interface and side 
walls are steeper. 
The grid sensitivity study shows that our grid resolution is sufficient to support the main conclusions in the paper.
}
\bibliographystyle{jfm}
\bibliography{Ref.bib}

\end{document}